\documentstyle[epsf,epsfig,12pt]{article}
\mark{{}{}}
\def\nll{ \nonumber \\}

\def\lb{\left(}
\def\rb{\right)}

\def\z0{Z}
\def\gf{G_{\mu}}
\def\zm{M_{_Z}}

\def\gev{{\hbox{GeV}}}

\def\hm{M_{_H}}
\def\wm{M_{_W}}

\def\fb{\overline f}
\def\barq{\overline q}
\def\bb{\overline b}
\def\bard{\overline d}

\def\barc{\overline c}
\def\bars{\overline s}
\def\baru{\overline u}

\def\bn{\overline{\nu}}

\def\i3f{I^{(3)}_f}

\def\osp2{16\,\pi^2}
\def\ap2{\left(p^2\right)}

\def\gev{{\hbox{GeV}}}

\def\s0h{\sigma^h_0}
\def\ba{\begin{eqnarray}}
\def\ea{\end{eqnarray}}

\def\beq{\begin{equation}}
\def\eeq{\end{equation}}
\def\bea{\begin{eqnarray}}
\def\eea{\end{eqnarray}}
\def\barr{\begin{array}}
\def\earr{\end{array}}
\def\bc{\begin{center}}
\def\ec{\end{center}}
\def\btab{\begin{tabular}}
\def\etab{\end{tabular}}

\begin{document}

\title{\bf STANDARD HIGGS BOSON\\
SEARCHES AT LEP~2}
\vskip 2cm
\author{
Giampiero PASSARINO$^{ab}$,
}
\date{}

\maketitle

\begin{itemize}

\item[$^a$]
             Dipartimento di Fisica Teorica,
             Universit\`a di Torino, Torino, Italy
\item[$^b$]
             INFN, Sezione di Torino, Torino, Italy

\end{itemize}

\vskip 1.5cm
The process $e^+e^- \to \bb b \fb f$ with $f = \nu,$ or $f =\,$ 
charged lepton or $f = u,d\,$ quark is analyzed in the range of LEP~2
energies. In the near future LEP~2 will represent a unique opportunity for a 
direct search of a Higgs boson of mass $> 65\,$GeV. The whole emphasis has
been put on a self consistent study of the standard Higgs boson properties.
Indeed there is the actual possibility of testing a calculation which uses
the full matrix elements for $e^+e^-$ annihilation into $4$-fermions
with the experimental data, beyond the usual approximation of computing 
production cross section $\times$ branching ratios of Higgs into decay 
products. Of course, given the fact that the number of collected events at 
LEP~2 will be limited and because of the low statistics one could proceed 
with a different strategy: for discovery physics elaborated tools are 
unnecessary. However the precise theoretical calculation is at our disposal 
already now, thanks to the combined effort of several groups, and it is 
extremely interesting and appealing to apply the 
corresponding machinery to an analysis -- as complete as possible -- of 
standard Higgs boson physics. Calculation of the total cross sections and of 
different kinds of differential distributions at the partonic level are made 
available to access the widest information for choosing cuts, for discussing 
the physics of a standard Higgs boson with a mass around $90\,$GeV where the 
Higgs and the $Z$ signal become degenerate and for a comprehensive analysis 
of the various background components. 
\vskip 1cm
\begin{center}
{Submitted to Nucl. Phys. B}
\end{center}

\newpage

\section{Introduction}

With the new experimental data from the LEP Collaborations and the increasing
achieved experimental accuracy there is an indirect evidence for a Higgs boson 
with a mass approximately below $500\,$GeV in the minimal standard 
model of the unified electroweak interactions~\cite{ichep}.

This is not the right place where to discuss the precision tests of the 
standard model and we only summarize few basic facts.
In principle we still have some minor problem in understanding the results of 
the fits to the experimental data, essentially we would like to understand 
when the $\chi^2(\hm)$ shape is unstable with respect to normal fluctuations 
of the experimental data in the large $\hm$ tail.
Also it appears that some clash between the LEP data and the SLD left-right 
asymmetry is still present, however for the first time in the LEP~1 history
the $\chi^2_{min}(\hm)$ has overcome some previous and unnatural tendency to 
be in the forbidden region, $\hm < 65\,$GeV, thus requiring the unnatural 
introduction of yet another penalty function. After the new results for 
$R_{b,c}$ the goodness of the fit has considerably improved upon the past and 
now the minimal standard model and the minimal supersymmetric standard model 
are, more or less, on equal footing in describing the experimental results. 

At any rate and roughly speaking one could make the following observations: it 
is still premature to give something more precise than an approximate
upper bound on the Higgs boson mass of $500\,$GeV at $95\%$ of CL.
Indeed as a result of the our fit~\cite{topaz0} we find
\vskip 0.5cm

\begin{eqnarray}
\hm &=& 143.5\,\gev  \nll
\hm &\leq& 431\,\gev \qquad {\hbox{at}} \qquad 95\% \quad {\hbox{CL}}.
\end{eqnarray}
\vskip 0.5cm

\noindent 
Our predictions for various quantities at the value of the Higgs boson mass
corresponding to the minimum of the $\chi^2$-distribution are shown in table 
$1$.

\begin{table}[hbtp]
\begin{center}
\begin{tabular}{|c|c|c|c|}
\hline
Observables  & Exp. & Theory & Comments\\
\hline
   &      &        & \\
$M_{_H}\,$(GeV) &  $> 65\,$GeV  & $143.5$ (fixed)   &  $< 431$  at $95\%$ CL    \\
   &      &        & \\
$\chi^2$ & &  $18.22/13$ & \\
   &      &        & \\
$m_t\,$(GeV)    &  $175 \pm 6$ & $172 \pm 5$ & penalty 
in the fit \\
   &      &        & \\
$\alpha^{-1}_{light}(M_{_Z})$  &  $128.89 \pm 0.09$ & $128.905 \pm 0.087$ 
&                \\
   &      &        & \\
$\alpha_s(M_{_Z})$  &  --  & $0.1194 \pm 0.0037$ 
& th. err. not included \\
   &      &        & \\
$m_b\,$(GeV) & $4.7 \pm 0.2$ & $4.68 \pm 0.24$  &  \\
   &      &        & \\
$R_l$ & $20.778 \pm 0.029$ & $20.754 \pm 0.025$ & \\
   &      &        & \\
$\sin^2\theta^e_{eff}$ & $0.23061 \pm 0.00047$ & $0.23159 \pm 0.00022$
 & \\
   &      &        & \\
$R_b$ & $0.2178 \pm 0.001144$ & $0.2158 \pm 0.0002$ & correlated \\
   &      &        & \\
$R_c$ & $0.1715 \pm 0.005594$ & $0.1723 \pm 0.0001$ &     "      \\
   &      &        & \\
$\wm\,$(GeV)   & $80.356 \pm 0.125$  & $80.350 \pm 0.031$  & \\
   &      &        & \\
$A^0_{_{FB}}(b)$ & $0.0979 \pm 0.0023$ & $0.1026 \pm 0.0012$ & \\
   &      &        & \\
   &      &        & \\
\hline
\end{tabular}
\end{center}
\label{tab1}
\caption{Theory versus Experiments -- a fit at the $Z$ resonance.}
\end{table}

\noindent
However with the indication from the LEP~1 and SLD data -- augmented with
$1 - \wm^2/\zm^2$ from $\nu N$-scattering, the $W$ mass from collider data
and $m_t$ from CDF and D0 -- and the advent of LEP~2 we should perhaps stop 
worrying about Tails$\&$Fits and we should instead start to understand how a 
Higgs boson - in the LEP~2 energy range - looks like in a real 
environment~cite{exp}. 
Indeed in the near future LEP~2 will be the only opportunity for a direct 
search of a Higgs boson of mass $> 65\,$GeV.

For the first time at LEP most of the events in the $e^+e^-$ annihilation
will have four fermion in the finals state and we have the unique opportunity
of testing a complete calculation with the experimental data, well beyond
the usual approximation of computing production cross section $\times$
branching ration of Higgs into decay products. Of course, given the fact that
the number of collected events at LEP~2 will be limited and because of the low
statistics one could proceed with a well defined strategy: for discovery 
physics sophisticated tools are unnecessary. However we do have precise 
theoretical calculations~\cite{yr2} at our disposal already now and it is 
extremely interesting and appealing to apply our machinery to an analysis 
-- as complete as possible -- of the Higgs physics.
With an increasing degree of complexity one goes through the following ladder 
of approximations

\begin{itemize}

\item[-] $e^+e^- \to Z^* H^*$, followed by the decays $Z^* \to \fb f$
and $H^* \to \bb b$. Under the assumption that the Higgs production at
LEP~2 is dominated by the Higgsstrahlung process $e^+e^- \to Z^* \to ZH$
the latter factorization is justified by the small Higgs width but the former
one is not good enough because of the much larger $Z$ width. Differential
distributions are not accessible. The additional diagrams leading to the same 
final state are not available.

\item[-] $e^+e^- \to \fb f H^*$, followed by the decay $H^* \to \bb b$. 
This works under the hypothesis that the c.m. energy is such that the fusion 
diagrams can be neglected. Again differential distributions are not accessible.
The additional diagrams leading to the same final state are not available.

\item[-] The full tree-level calculation $e^+e^- \to \bb b \fb f$.
No approximation is made, differential distributions are available and the
background is under control (with some limitations to be discussed below).

\end{itemize}

\noindent
For a complete discussion of the Higgs boson branching ratios, inclusive
of their radiative corrections, we refer to~\cite{sh} and to references
therein.

First of all we have learned how to deal with unstable particles 
in a fully satisfactory field theoretical context~\cite{up}, thus the 
properties of the Higgs boson at LEP~2 can and should be inferred from the 
analysis of the following processes:
\vskip 0.5cm

\begin{eqnarray}
e^+e^- &\to& \bb b \mu^+\mu^-,\qquad \bb b e^+e^-,\nll
{}&& \bb b \bn \nu,\qquad \bb b \baru u,(\barc c),\nll
{}&& \bb b \bard d,(\bars s), \qquad \bb b \bb b.
\end{eqnarray}
\vskip 0.5cm

\noindent
with all the complications arising from flavor mis-identification. Thus the
three typical signatures are

\begin{enumerate}

\item two jets + a charged lepton pair,

\item two jets + missing energy and momentum,

\item four jets

\end{enumerate}

\noindent
Thus our aim is to provide predictions for these processes at the partonic 
level, giving useful informations on the strategy for Higgs searches and
on the relative importance of the background to the Higgs signal.
The ideal procedure would be to analyze all the above channels with some event
generator - the ultimate one - which should

\begin{itemize}

\item[-] account for the experimental setup, optimized for the search of
the Higgs boson,

\item[-] include a self-consistent set of radiative corrections~\cite{yr2ww}, 
to optimize the theoretical accuracy,

\item[-] be interfaced with standard hadronization packages~\cite{tb}.

\end{itemize}

A broad separation can be set between the two alternative poles:
a dedicated electroweak calculations or a general purpose simulations.
In this context a crucial role is played by the hadronization process~\cite{tb}.

At least in first approximation we could say that a dedicated electroweak 
calculation will describe the electroweak content of the processes to very 
high accuracy but it will lack perturbative parton shower or non-perturbative
hadronization. This raises the question of its reliability for the study
of hadronic or mixed hadronic-leptonic final states. Even though a pragmatic
solution, adopted by many authors, consists in standard interfacing with 
parton-shower and hadronization programs we still insist on the importance 
of presenting the most precise predictions for cross sections and 
distributions with the inclusion of final states QCD perturbative corrections. 

It has been shown by the LEP~1 collaborations
that such predictions are indeed of the upmost importance for understanding
the underlying physical properties of the model once the proper
de-convolution procedure is applied to the data. For this reason we are
still thinking that a correct (theoretical) treatment of the problem at
the level of exact and full matrix elements (including perturbative QCD
corrections) will be essential in understanding several features, not
least the quantitative effect of some of the most common approximations
and the relevance of background versus signal, all of this in presence
of some set of simple but realistic enough cuts.

Therefore we want to discuss the Higgs boson properties at LEP~2 by
including all diagrams for all given channels at the $0.1\%$ level of 
technical precision (or better) and by including the best available set of 
corrections, i.e. initial state QED radiation through the structure function
approach~\cite{sf}, running quark masses~\cite{rqm}, naive QCD (NQCD) final 
state corrections. 

When using the exact matrix elements it becomes desirable to include final 
state QCD corrections, even when kinematical cuts are imposed on the outgoing
fermions. Lacking a complete calculation we have adopted in this paper the so 
called naive approach where by naive QCD we mean a simple recipe where the 
total  $H(Z)$-width and the cross section are corrected by some simple 
multiplicative factors.
This naive approach, consequence of our ignorance about the complete
result, would be correct only for the double-resonant diagrams within a 
fully extrapolated setup. 

One final comment is in order here on the general philosophy of our 
approach, which is shared by many other authors already present in the
literature.
To a large extent the standard Higgs boson -- being such a narrow resonance --
could be considered as an almost stable particle and one could be tempted 
and motivated in investigating independently its production mechanism and its 
decay processes. One more requirement has to be fulfilled in this respect,
the Higgs boson physics must be something to be added incoherently to the 
background. As already noticed in the massless limit signal and background add
incoherently but there are production mechanisms, say for $e^+e^- \to \bb b
\bn_e \nu_e$, where different contributions have an interference,
typically the fusion process with the Higgsstrahlung one. Since $\bb b +\,$
neutrinos represent $20\%$ of the signal at LEP~2 energies this has to be 
correctly taken into account as described elsewhere in this paper with
the global description of the Higgs-fusion relevance as a function of the
kinematical cuts. 

To understand the general degree of complexity of the calculation based on
the full matrix element approach we have shown in table $2$ a simple counting 
of the diagrams which occur in all processes.

\begin{table}[hbtp]
\begin{center}
\begin{tabular}{|c|c|}
\hline
Final state  & Class/max $\#$ of diagrams\\
             &                           \\
\hline
             &                           \\
$\bb b \mu^+\mu^- $  & NC25 \\
             &                           \\
$\bb b e^+e^- $  & NC50\\
             &                           \\
$\bb b \bn_{\mu} \nu_{\mu} $  & NC25\\
             &                           \\
$\bb b \bn_e \nu_e $  & NC21\\
             &                           \\
$\bb b \baru u,(\barc c) $  & NC33\\
             &                           \\
$\bb b \bard d,(\bars s) $  & NC33\\
             &                           \\
$\bb b \bb b $  & NC68 \\
             &                           \\
\hline
\end{tabular}
\end{center}
\label{tab2}
\caption{Diagrams required (including gluons) for the process $e^+e^- \to
\bb b \fb f$.}
\end{table}

\noindent
Our terminology is an obvious extension of the one introduced in~\cite{yr2}.
Thus NC-type refers to the production of two fermion-antifermion pairs
$(\fb_i, f_i) + (\fb_j, f_j)$ where $i,j$ are generation indices. 
To give an example NC50 is the natural extension of NC48 with two extra
diagrams corresponding to Higgstrahlung and $ZZ$ fusion.
Here all the processes are of the Neutral Current type (NC) and in
$\bb b \bb b$ we have neglected those diagrams which correspond to a 
$\bb b$ pair production from a $b(\bb)$ leg.
Moreover in the massless limit the interference between the Higgs signal and 
its background is zero, fact which greatly simplifies the calculation.
This is a consequence of the coupling of massless fermions to either 
spin-vectors or to the Higgs, a spin-scalar.

Thus the strategy for the calculation is relatively simple.
There are already several examples of Higgs studies in the literature
which have addressed and solved several important issue. 
It is rather difficult to summarize all the work done in the recent past. 
Tentatively we point out the relevance of the pioneering work of
E.~Boos and M.~Dubinin~\cite{bd} on the Monte Carlo calculation of the 
processes
\vskip 0.5cm

\begin{eqnarray}
e^+e^- \to Z \bb b, \nll
e^+e^- \to \bb b \mu^+ \mu^-.
\end{eqnarray}
\vskip 0.5cm

\noindent
The same process has been extensively analyzed by G.~Montagna, O.~Nicrosini 
and F.~Piccinini~\cite{cod5}.
Also we stress the importance of the semi-analytical approach to Higgs
production at LEP~2, as illustrated by the work of D.~Bardin, A.~Leike and
T.~Riemann~\cite{dz}. 
Important contributions to understand Higgs physics - even though not from
the point of view of LEP~2 energy range -- can be found in the work of
P.~Grosse-Wiesmann, D.~Haidt and J.~Schreiber~\cite{ob1} and in the one of 
P.~Janot~\cite{ob2}.

A complete and exhaustive comparison for Higgs physics can be found in the
Report of the Workshop in Physics at LEP~2, where the signal and background
cross sections for the process $e^+e^- \to \bb b \mu^+ \mu^-$ have been
compared among the following FORTRAN codes: CompHEP~\cite{cod1}, 
EXCALIBUR~\cite{cod2}, FERMISV~\cite{cod3},GENTLE/4fan~\cite{cod4}, 
HIGGSPV~\cite{cod5}, HZHA/PYTHIA~\cite{cod6}, WPHACT~\cite{cod7} and 
WTO~\cite{cod8}. It should be stressed that during this comparisons no
QCD corrections have been applied and the $b$-quark mass has been fixed
to its pole value. In this respect our results represent another step
towards the ultimate Higgs prediction.

Other codes designated for $4$-fermion physics have been quite active
recently. In the previous list we have only quoted the participants in
the Higgs physics working group at the LEP~2 workshop who have actually 
produced results for the published tables. For further general references
one should consult ref.~\cite{to}.

Among the most recent analyses we would like to quote the work of
Katsanevas and collaborators~\cite{kat} who present a study of $e^+e^- \to 
\bb b\bn \nu$ at c.m. energies $150 \leq \sqrt{s}$(GeV)$ \leq 240$ and 
where the important differential distributions for the Higgs boson and the 
background components are studied, providing information useful for choosing 
cuts in Higgs searches.

In this paper we present a modest contribution to the discussion by analyzing 
many different final states which are important for the discovery of a minimal
standard model Higgs boson at LEP~2. This we will do in one study, without 
selecting any particular signature and by using the FORTRAN code 
WTO~\cite{cod8}. 

Although the issue of the fermion masses
will be addressed later in the paper here we stress that all the reported 
results are computed with massless fermions, Yukawa couplings excluded.
Masses are not a limitation of principle in the formalism upon which WTO
is based but they are not yet implemented, thus we refer to those code which
have already produced results with $m_b \not= 0$, noticeably 
CompHEP~\cite{cod1}, and WPHACT~\cite{cod7}. Also we will devote little space 
to $\bb b \bb b$ final states and signal where, to the best of our knowledge, 
the most reliable results are achievable with WPHACT.

The outline of the paper is a s follows. In sect 2 we present and discuss in 
details our strategy for the calculation. In sect. 3 we briefly discuss the
theoretical uncertainties associated with the formulation of the problem and 
with the choice of the input parameters. The presentation of all processes,
their background and of the set of kinematical cuts is in sect. 4 while
a detailed discussion of the numerical results is contained in sect. 5.
Our conclusions are shown in sect. 6.

\section{Strategy of the calculation}

In this section we will show how the calculation of the Higgs boson signal and
of its background is organized. At the same time the feasibility of the 
approach will be illustrated. All fermions masses which occur in
\vskip 0.5cm

\begin{equation}
e^+e^- \to \bb b \fb f,
\end{equation}
\vskip 0.5cm

\noindent
are neglected but for the $b$-quark mass in the Yukawa coupling and for the 
$b$-quark, $c$-quark and $\tau$ masses in the decay width. Quark masses are 
running and evaluated according to~\cite{rqm}
\vskip 0.5cm

\begin{eqnarray}
{\bar m}(s) &=& {\bar m}(m^2)\,\exp\left\{
-\,\int_{a_s(m^2)}^{a_s(s)}\,dx \frac{\gamma_m(x)}{\beta(x)}\right\} \;,
\nll
m &=& {\bar m}(m^2)\,\left[ 1 + \frac{4}{3} a_s(m) + K a_s^2(m)\right]\;,
\end{eqnarray}
\vskip 0.5cm

\noindent
where $m = m_{pole}$ and $K_b \approx 12.4, K_c \approx 13.3$.

To summarize we have neglected the fermion masses in the phase space while 
keeping them in the Yukawa couplings. The effect of including the masses in 
some effective approximation will be discussed later in the paper.
The Higgs width is computed with the inclusion of the $H \to gg$ 
channel. 
The most complete treatment will therefore start with some input value for
$\alpha_s(\wm)$ and it will evolve $\alpha_s$ to the scale $\mu = \hm$, 
will evaluate the running $b,c$-quark masses and finally compute
\vskip 0.5cm

\begin{eqnarray}
\Gamma_{_H} &=& {{G_G\hm}\over {4\,\pi}}\,\left\{ 3\,\left[ 
m_b^2(\hm) + m_c^2(\hm)\right]\,\left[ 1 + 5.67\,{\alpha_s\over \pi}
+ 42.74\,\lb{\alpha_s\over \pi}\rb^2\right] + m_{\tau}^2\right\} \nll
{}&+& \Gamma_{gg},  \nll
\Gamma_{gg} &=& {{\gf\hm^3}\over {36\,\pi}}\,{{\alpha_s^2}\over {\pi^2}}\,
\lb 1 + 17.91667\,{\alpha_s\over\pi}\rb.
\end{eqnarray}
\vskip 0.5cm

\noindent
As already indicated NQCD is included by evoluting $\alpha_s(\wm)$(input) to 
$\alpha_s(\hm)$ and the Higgs boson signal is multiplied by
\vskip 0.5cm

\begin{eqnarray}
\delta_{_{QCD}} &=& 1 + 5.67\,{{\alpha_s}\over {\pi}} + 
42.74\,\lb {{\alpha_s}\over {\pi}}\rb^2,  \nll
\alpha_s &=& \alpha_s(\hm).
\end{eqnarray}
\vskip 0.5cm

\noindent
Similarly whenever we consider $e^+e^- \to \bb b \barq q$ there will be an 
additional NQCD correction factor $1 + \alpha_s/\pi$, where $\alpha_s$ is now
evaluated at a scale $\mu = \zm$.

Admittedly this is not the most satisfactory solution to the problem of
final state QCD corrections. Besides what we have already indicated it must 
be said that several QCD diagrams (gluon-exchange) are, in this way, neglected.
For instance there are multi-leg diagrams, including boxes or even higher.
In other words NQCD is equivalent to shrink the whole electroweak part
of a diagram to a point and to apply QCD radiation to each $\barq_i q_i$ 
pair at a fixed scale -- $\mu = \hm$ for $\bar b b$ , $\mu = \zm$ for the 
rest -- while neglecting at the same time all kinematical cuts.
To illustrate the expected properties we have shown in table $3$ some of the 
parameters of the Higgs boson for different values of its mass:

\begin{table}[hbtp]
\begin{center}
\begin{tabular}{|c|c|c|c|}
\hline
Parameter  & $\hm = 80\,$GeV & $\hm = 90\,$GeV & $\hm = 100\,$GeV\\
\hline
           &                 &                 &                 \\
$\Gamma_{_H}$   & $1.8515\,$MeV  & $2.0601\,$MeV & $2.2734\,$MeV \\
           &                 &                 &                 \\
$m_b(\hm)$      & $2.731\,$GeV   & $2.702\,$GeV  & $2.676\,$GeV  \\
           &                 &                 &                 \\
$m_c(\hm)$      & $0.553\,$GeV   & $0.547\,$GeV  & $0.542\,$GeV  \\
           &                 &                 &                 \\
$\alpha_s(\hm)$ & $0.12557$      & $0.12323$     & $0.12121$     \\
           &                 &                 &                 \\
\hline
\end{tabular}
\end{center}
\label{tab3}
\caption{Input is $\alpha_s(\zm)= 0.123$, $m_b = 4.7\,$GeV and 
$m_c =1.55\,$GeV.}
\end{table}

To continue our description of the chosen strategy we will say that the matrix 
elements are generated in our calculation through the 
helicity formalism of ref.~\cite{mhf} and they are compact expressions 
completely given in terms of the invariants which describe the process. Also 
the momenta of the final states are, component by component, given in terms
of the invariants used in the integration over the phase space, thus allowing
to implement the kinematical cuts with an analytical control.

We adopted this procedure to compute all the relevant cross sections with
the FORTRAN code WTO{\footnote{A preliminary collection of results from WTO for
Higgs searches can be found in ref.~\cite{crad}}}. However for a Higgs study 
all kind of distributions at
the parton level are extremely important. For this reason we have extended the
original version of WTO in order to allow for the generation of unweigthed
events and for the storage of their four-momenta. After that all kind of
distributions can be analyzed, according to a well established procedure.
The full description of WTO 2.0 and of the methods adopted to generate
unweigthed events will be given elsewhere~\cite{wto2}.
For a large number of differential distributions therefore WTO can work
under two alternative strategy, a deterministic integration with analytical
control over the boundaries of the phase space or an event generator.
In the former case a differential distribution can be generated bin by bin
with a slow but very precise procedure while in the latter the same 
distribution is generated in one run with a fast procedure. In a large number
of cases we have therefore confronted the results of the two approaches with 
a quite satisfactory agreement.

Whenever computing processes with a Higgs boson exchange one is usually faced 
with the problem of including the fermion masses or not in the calculation.
The Yukawa coupling
\vskip 0.5cm

\begin{equation}
\fb H f \qquad -\,\frac{1}{2}\,g\,{{m_f}\over M},
\end{equation}
\vskip 0.5cm

\noindent 
are of course there but usually the fermion mass effects in the rest of the 
matrix element and in the phase space are neglected. The main reason for
doing that has always to do with the CPU time needed for the calculation and,
usually is never a matter of principle. In order to understand the
corresponding behavior we have investigate an effective mechanism for
introducing fermion mass effects. Since the Higgs width is extremely
narrow and since the running masses are at most of the order of $2\div 3\,$
GeV we can think of including mass effects in a narrow width approximation.
For instance in the $ZH$ production mechanism and for some fully extrapolated
setup we can write the basic off-shell $ZH$ production as
\vskip 0.5cm

\begin{eqnarray}
\sigma_{_{ZH}}(s) &=& \int_0^s\,ds_{_1}\,\rho_{_H}(s_{_1})\,
\int_0^{(\sqrt{s} - \sqrt{s_{_1}})^2}\,ds_{_2}\,\rho_{_Z}(s_{_2})
\sigma_{_{ZH}}\left(s;s_{_1},s_{_2}\right),  \nll
\rho_{_V}(s_i) &=& \frac{1}{\pi}\,{{\sqrt{s_i}\,\Gamma_{_V}(s_i)\times
BR(i)}\over {\mid s_i - M_{_V}^2 +i\,\sqrt{s_i}\,\Gamma_{_V}(s_i)\mid^2}}.
\end{eqnarray}
\vskip 0.5cm

\noindent 
In narrow width approximation we multiply the $e^+e^- \to \bb b \fb f$ 
cross section by a factor $F_{mass}$ given by
\vskip 0.5cm

\begin{eqnarray}
F_{mass} &=& \beta^3\,\left(1 + {{\Gamma_b}\over {\Gamma_R}}\right)\,
\left(1 + \beta^3\,{{\Gamma_b}\over {\Gamma_R}}\right)^{-1},  \nll
\beta^2 &=& 1 - 4\,{{m^2_b(\hm)}\over {\hm^2}},  \nll
\Gamma_R &=& \Gamma_H - \Gamma_b.
\label{emass}
\end{eqnarray}
\vskip 0.5cm

\noindent
The numerical impact will be discussed in  the section about numerical results.
In general however our calculation is exact at ${\cal O}(m^2_b(\hm))$ and
for this reason in the process $e^+e^- \to \bb b \bb b$ we have neglected 
diagrams which correspond to a $\bb b$ pair produced by a $b(\bb)$ line
since they give contributions to the cross section of ${\cal O}(m^4_b(\hm))$.

\section{Theoretical uncertainties}

Each calculation aimed to provide some estimate for $4$-fermion production
is, at least nominally, a tree level calculation. Among other things it will
require the choice of some set of input parameters and of certain relations 
among them. This is usually referred in the literature, although improperly, 
as the choice of the Renormalization Scheme (RS).

So far an attempt to investigate the problem has been performed by GENTLE/4fan
and by WTO~\cite{yr2}. More recently the size of the theoretical uncertainties 
for neutral current processes has been addressed by WPHACT~\cite{pc} and WTO.

There are several sources for the theoretical uncertainty and
no fully reliable estimate of the theoretical error can be given.
At most we can produce a rough estimate by applying few options 
connected with the choices of the Renormalization Scheme. 

Typically we have at our disposal four experimental data point 
(plus $\alpha_s$), i.e. the measured vector boson masses $\zm, \wm$ and the
coupling constants, $\gf$ and $\alpha$. However we only 
have three bare parameters
at our disposal, the charged vector boson mass, the $SU(2)$ coupling constant
and the sinus of the weak mixing angle. While the inclusion of one loop
corrections would allow us to fix at least the value of the top quark mass
from a consistency relation this cannot be done at the tree level. Thus 
different choices of the basic relations among the input parameters can lead 
to different results with deviations which, in some case, can be sizeable.

For instance we have considered the Higgs background ($\hm = \infty$) at one 
particular energy, $190\,$GeV, and computed the corresponding cross sections
in two among the most popular schemes. This background is therefore
affected by the theoretical error shown in table $4$.

\begin{table}[hbtp]
\begin{center}
\begin{tabular}{|c|c|}
\hline
Process  & 1-($G_F$ scheme)/($\alpha$ scheme) (permill)\\
\hline
         &                                             \\
$\bb b \bn_{\mu} \nu_{\mu}$ &  $  0.86 $\\
         &                                             \\
$\bb b \mu^+ \mu^-           $ &  $  2.23 $\\
         &                                             \\
$\bb b \nu_e \nu_e           $ &  $  2.44 $\\
         &                                             \\
$\bb b e^+ e^-               $ &  $  8.05 $\\
         &                                             \\
$\bb b \baru u               $ &  $ -3.21 $\\
         &                                             \\
$\bb b \bard d               $ &  $ -3.03 $\\
         &                                             \\
\hline
\end{tabular}
\end{center}
\label{tab4}
\caption{Differences (in permill) induced by different Renormalization Schemes 
for the Higgs background at $190\,$GeV.}
\end{table}

\noindent
Roughly speaking we can say that the theoretical uncertainty associated with 
the choice of the RS is most severe whenever low-$q^2$ photons dominate,
both for $q^2 >0$ and $q^2 < 0$. Indeed we have for the most popular choices

\begin{itemize}

\item[-] The $\alpha(\zm)$ scheme.
\vskip 0.5cm

\begin{equation}
s_{_W}^2 = {{\pi\alpha}\over {{\sqrt 2}\gf\wm^2}},  \qquad 
g^2 = {{4\pi\alpha}\over {s_{_W}^2}}
\end{equation}
\vskip 0.5cm

\item[-] The $\gf$ scheme
\vskip 0.5cm

\begin{equation}
s_{_W}^2 = 1 - {{\wm^2}\over {\zm^2}}, \qquad
g^2 = 4{\sqrt 2}\gf\wm^2. 
\end{equation}
\vskip 0.5cm

\end{itemize}

Thus in the $\gf$-scheme the e.m. coupling is governed by $\alpha =
1/131.22$ while in the $\alpha(\zm)$-scheme it is $\alpha = 1/128.89$
which accounts for a $2\%$ difference. This will propagate into approximately
a $10\%$ difference between the two schemes at low-$q^2$ for diagrams with 
two photons. Processes with both time-like or space-like photons are
therefore severely affected unless protective cuts are imposed.
For this reason the effect is larger in $e^+e^- \to \bb b e^+ e^-$, due to
the presence of a multi-peripheral diagram with photons in the $t$-channel.

Essentially we may distinguish between $s$-channel photons where the difference
between the two schemes can be made arbitrarily negligible by cutting on the
corresponding $\gamma^* \to \fb f$ invariant mass and $t$-channel photons
where one would have to impose more stringent cuts on the corresponding
scattering angle.

Actually there is a third alternative, somehow dictated by the LEP~1
framework which in some case could be more relevant. First we compute
the running of $\alpha$ up to a scale $\mu = \zm$. This can be done by
including the leptons and the top quark perturbatively while the light
quarks are accomodate through dispersion relations~\cite{jeg}. Next we define
\vskip 0.5cm

\begin{eqnarray}
s_{_W}^2 &=& \frac{1}{2}\,\left[1 - \sqrt{1 - {{4\,\pi\alpha(\zm)}\over
{\sqrt{2}\,\gf\zm^2}}}\right]] \nll
g^2 &=& 4{\sqrt 2}\gf\zm^2\,c_{_W}^2. 
\end{eqnarray}
\vskip 0.5cm

\noindent
Other additional sources of uncertainty are in the parametrization of the 
QED structure functions~\cite{sf}, in the treatment of the scale $\mu$ in the 
QCD corrections, 
expecially so for the scale of $\alpha_s$ in the NQCD correction factor. The 
default established during the LEP~2 workshop usually consists in inserting 
$\alpha_s$ at fixed $\mu$ even for internal gluons in a process like $e^+e^- 
\to \bb b \barq q$. In this case the scale is generally chosen to be $\zm$ or
$\hm$. A better choice could be to use $\alpha_s$ evaluated at a running 
virtuality, i.e. $\alpha_s({\hat s})$ where for instance ${\hat s}$ is the 
invariant mass of the $\bar q q$ pair.
With this choice however some lower cuts on the invariant masses are required 
to avoid the non-perturbative, low-$q^2$, regime.

On top of the theoretical uncertainties there are additional problems,
flavor mis-identification and correct treatment of the background. 
Experimentally one must extract 
the Higgs signal from all final states consisting of a pair of (imperfectly)
$b$-tagged jets + remaining products (including the missing ones).
The probabilities of a light quark, a $c$-quark or a $b$-quark
jet to be confused with a $b$-quark are non zero. The effect of flavor
mis-identification modifies the original branching ratios.
For instance a 2-jet final state $q_{_1}\barq_{_2}$ with given quark flavors
can be characterized by a two-by-two matrix~\cite{ob1}
\vskip 0.5cm

\[\left(\begin{array}{cc}
(1-P_{1b})\,(1-P_{2b}) & (1-P_{1b})\,P_{2b} \\
P_{1b}\,(1-P_{2b})     & P_{1b}\,P_{2b} \\
\end{array}\right)\]
\vskip 0.5cm

\noindent
In order to introduce a discussion for the Higgs background we observe that
at LEP~2 a large fraction of Higgs events will be of the type
$\bb b \bn \nu $ ($\approx 20\%$). There are potentially large 
backgrounds in the process
$e \nu_e c s$ with flavor mis-identification and with the electron lost in the
beam-pipe. A safe estimate requires including $m_e$ in the calculation since
we go down to $\theta_e = 0$ -- zero scattering angle for the electron --
where moreover gauge invariance is in danger.
Another example is given by $l^+ l^- \bb b$ with the leptons lost in the 
beam-pipe. Again it requires a finite lepton mass because of divergent 
multi-peripheral diagrams. No reliable estimate has been given so far in
the literature.

To explain in more details the gauge invariance issue we can say that
here we are dealing with CC20 diagrams
with $t$-channel photons which induce an apparent singularity at zero
scattering angle. This is of course cured by avoiding the approximation
of massless fermions but there is more. 
Any calculation for $e^+e^- \to 4$-fermions is only nominally a {\it tree
level} approximation because of the presence of charged and neutral, unstable 
vector bosons and of their interaction with photons.

Unstable particles require
a special care and their propagators, in some channels, must necessarily
include an imaginary part or in other words the corresponding $S$-matrix
elements will show poles shifted into the complex plane. In any 
field-theoretical approach these imaginary parts are obtained by performing
the proper Dyson resummation of the relative two-point functions, which
at certain thresholds will develop the requested imaginary component.
The correct recipe seems representable by a Dyson resummation of fermionic 
self-energies where only the imaginary parts are actually included. As a 
result the vector boson propagators will be inserted into the corresponding 
tree level amplitudes with a $p^2$-dependent width. Its has already been 
noticed by several authors \cite{up} that even this simple idea gives 
rise to a series 
of inconsistencies, which sometimes may give results completely inconsistent 
even from a numerical point. The fact is that the mere introduction of a 
width into 
the propagators will inevitably result, in some cases, into a breakdown of 
the relevant Ward identities of the theory with a consequent violation of some 
well understood cancellation mechanism. In the CC20 case the effect of 
spoiling a cancellation among diagrams results into a numerical catastrophe.
The solution of this apparent puzzle is by now well know and amounts to the 
inclusion of the so-called Fermion-Loop scheme~\cite{up}. A reliable estimate 
of this background would require the introduction of both a finite electron
mass and of the Fermion-Loop scheme.
A full description of The Fermion Loop scheme is well beyond the scope of
our paper. There are two versions of the scheme, one where roughly speaking
one adds the imaginary parts of all fermionic one loop diagrams and a
second more complete one which amounts to the inclusion of the full
${\cal O}(\alpha)$ fermionic corrections, inclusive of the proper
treatment of the vector boson complex poles.
\section{The Higgs signal and its background}

In order to analyze the Higgs signal versus background at LEP~2 we fix
our set of quasi-realistic cuts for the processes to be considered by WTO. At 
the parton level they are

\begin{enumerate}

\item $M(\bb b) \geq 50\,$GeV, $\mid M(\fb f) - \zm\mid \leq 25\,$GeV.
The former is to suppress the photon mediated $\bb b$ production
-- which decreases for larger $\sqrt{s}$. The latter reduces all contributions
which give a broad $M(\fb f)$ spectrum.

\item Lepton momenta $\geq 10\,$GeV.

\item Quark energies, $E_q \geq 3\,$GeV.

\item Lepton polar angles with the beams $\geq 15^o$.

\item For processes with neutrinos the angle of both b's with the beams
$\geq 20^o$ or of at least one b.

\item $\theta(l,q) \geq 5^o$.

\end{enumerate}

\noindent
This set of kinematical cuts is the one chosen during the last LEP~2
workshop and after that it has been termed Canonical Cuts (or CC in shorts).
For the $\bb b \bb b$ final state some additional selection will be
introduced in the next section.

We have already pointed out that our prediction are at the parton level
and moreover any realistic analysis will require several further acceptance
criteria, $b$-tagging and constrained morphology. Detecting a Higgs boson
at LEP~2 requires also the isolation of the signal from many different sources
of electroweak background. They have been classified recently in ref.~\cite{kat}
and here we would like to spend few more words of comment.

\begin{itemize}

\item[-] $s$-channel production of $\bb b$ jets with soft and undetected 
hadrons. This is a longstanding problem of the separation between $4$-fermion
final states from what one should really consider as radiative corrections
-- initial state pair production -- to $2$-fermion final states at LEP~2.
Here any calculation requires experimental guidance on the set of cuts 
needed to distinguish the two regimes. In principle there should be little
problem in interfacing $2$-fermion codes with $4$-fermion codes in order
to give a correct treatment of the relevant physical processes at LEP~2.
{\footnote{The interfacing of TOPAZ0($2$f) with WTO($4$f) is currently
under investigation}}
\item[-] $e^+e^- \to c s  \nu \tau(\to \nu \bn +\,$soft charged particles).
This is a typical CC10 process which is well under control even though the
existing dedicated electroweak codes should be interfaced with some
$\tau$-decay library.

\item[-] $e^+e^- \to e \nu_e c s$ with flavor mis-identification and the 
$e$ lost in the beam-pipe or $e^+e^- \to l^+ l^- \bb b$ with the leptons 
lost in the beam-pipe. We have already indicated that no reliable estimate
has been performed so far. Take for instance the typical CC20 process
$e^+e^- \to e \nu_e c s$, here gauge invariance becomes essential in the 
region of phase space where the scattering angle of the electron is small
and moreover in the same region the photon propagator behaves like
$1/m_e^2$. Thus both finite electron mass and $U(1)$ gauge invariance are
required for a meaningful cross section with the outgoing electron contained
in a small cone around the beam axis.

\end{itemize}

\section{Numerical Results and Comments}

To understand the Higgs boson search at LEP~2 through the subprocess 
contributions to the cross section as functions of the c.m. energy we have
reported in figure $1$ the cross sections for $e^+e^- \to
\bb b \fb f$ where $f = \mu,\nu_{\mu} (HZ$-component), $f = \nu_e (HZ +
WW$-components) and $f = e (HZ + ZZ$-components). 

In this figure we have shown
both the complete cross sections as well as the differences $\sigma - 
\sigma_{bckg}$ (where $bckg$ indicates the same process but with $\hm = 
\infty$) which illustrate how the $ZZ$ component is much less relevant
than the $WW$ one, which in turns is not negligible over the whole LEP~2
energy range. The relative importance of the $WW$-component, below the $HZ$
threshold, around it and slightly above is illustrated
in figure $2$ where we have compared $3\,\sigma(e^+e^- \to \bb b
\bn_{\mu} \nu_{\mu})$ with $3\,\sigma(e^+e^- \to \bb b \bn_{\mu} \nu_{\mu})
+ \sigma(e^+e^- \to \bb b \bn_e \nu_e)$.

To continue the discussion of our results we start from the evaluation of 
the cross sections as a function of $\sqrt{s}$ for $\hm = 80\,\gev, 90\,\gev, 
100\,\gev$ and $\infty$. They are show in figure $3$.
Whenever a process is indicated it has to be understood that the reference 
kinematical cuts, as described in the previous section, are applied. Our 
findings confirm the rule of thumb
\vskip 0.5cm

\begin{center}
{$\hm \approx \sqrt{s} - 100\,\gev$ for LEP~2 feasibility}
\end{center}
\vskip 0.5cm

\noindent
It is indeed evident that for the cross sections the ratio signal/background, 
for fixed $\hm$, has a maximum at $\sqrt{s} \approx \hm + 100\,$GeV.
If the Higgs is above $80\,GeV$ then the cross section is too small at 
$\sqrt{s} = 175\,$GeV to allow for a Higgs discovery, thus the $\sqrt{s} = 
190\,$GeV phase of the collider -- or a higher one -- will be needed.
Moreover at $\sqrt{s} \geq 190\,$GeV the $ZZ$ background is not negligible
and here is where a dedicated EW calculation becomes useful.
For instance we assume a Higgs mass of $100\,$GeV and compare
$\sigma$(signal) versus $\sigma$(bakground) at $195\,$GeV. The effect of the 
background is clearly shown in table $5$ to be of the order of $50\%$.

\begin{table}[hbtp]
\begin{center}
\begin{tabular}{|c|c|c|}
\hline
Process  & $\sigma$(signal) & $\sigma$(bakground) \\
\hline
                                 &         &        \\
$\bb b \bn_{\mu} \nu_{\mu}$ & 24.957  & 16.439 \\
                                 &         &        \\
$\bb b \mu^+ \mu^-           $ & 13.515  & 9.205  \\
                                 &         &        \\
$\bb b \nu_e \nu_e           $ & 29.068  & 16.874 \\
                                 &         &        \\
$\bb b e^+ e^-               $ & 15.631  & 11.449 \\
                                 &         &        \\
$\bb b \baru u               $ & 51.110  & 34.276 \\
                                 &         &        \\
$\bb b \bard d               $ & 63.924  & 42.277 \\
                                 &         &        \\
\hline
\end{tabular}
\end{center}
\label{tab5}
\caption{Signal versus Background cross sections in fb. The c.m. energy
is $195\,$GeV.}
\end{table}

\noindent
Another point where a dedicated electroweak calculation indeed makes 
substantial improvement is given by the process $e^+e^- \to \bb b \bn_e \nu_e$
process because of the presence of $t-$channel diagrams. It has already
been reported in ref.~\cite{yr2} that for $\sqrt{s} = 175\,$GeV 
and -- for instance $\hm = 90\,$GeV -- the difference between HZHA and the
average among HIGGSPV, WPHACT and WTO is approximately $42\%$, in a situation 
where the agreement between the dedicated Higgs codes is systematically 
better than $1\%$. Having or not control over the full set of diagrams clearly
makes a difference. 

A word of comment is needed at this point, it is
important to point out that the impact of the discrepancies is minimal on
the discovery potential of LEP~2. However when we come to the question of
the extraction of Higgs properties then the perspective changes, control
on the exact matrix elements is required to eliminate the largest discrepancies
and the effect of additional $\%$ level uncertainties -- of the order of
those coming from higher order corrections -- will require further work
to be fully understood.

Coming back to the cross sections they will be shown later on in several 
figures but -- for further reference and for comparisons -- we have also 
shown a reduced sample of results in table $6$.

\begin{table}[hbtp]
\begin{center}
\begin{tabular}{|c|c|c|c|c|}
\hline
Process/$\hm\,$(GeV)  & 80 & 90 & 100 & $\infty$ \\
\hline
\hline
$\sqrt{s} = 175\,$GeV            &         &        &        &        \\
\hline
                                 &         &        &        &        \\
$\bb b \bn_{\mu} \nu_{\mu}$ & 17.519(1) & 1.5849(8) & 0.9905(8) & 0.9082(8) \\
                                 &         &        &        &        \\
$\bb b \mu^+ \mu^-        $ &  9.243(1) & 1.1291(8) & 0.8265(8) & 0.7851(7) \\
                                 &         &        &        &        \\
$\bb b \nu_e \nu_e        $ & 22.570(2) & 3.3981(8) & 1.2797(8) & 0.7489(8) \\
                                 &         &        &        &        \\
$\bb b e^+ e^-            $ & 10.047(2) & 2.120(2)  & 1.870(2)  & 1.832(2) \\
                                 &         &        &        &        \\
$\bb b \baru u            $ & 36.30(1)  & 4.364(5)  & 3.171(4)  & 3.010(4) \\
                                 &         &        &        &        \\
$\bb b \bard d            $ & 45.44(1)  & 4.412(3)  & 2.878(3)  & 2.677(3) \\
                                 &         &        &        &        \\
\hline
\hline
$\sqrt{s} = 190\,$GeV            &         &        &        &        \\
\hline
                                 &         &        &        &        \\
$\bb b \bn_{\mu} \nu_{\mu}$ & 42.154(4) & 30.430(2) & 14.9913(6) & - \\
                                 &         &        &        &        \\
$\bb b \mu^+ \mu^-        $ & 22.339(3) & 16.213(1) &  8.330(1)  & - \\
                                 &         &        &        &        \\
$\bb b \nu_e \nu_e        $ & 47.062(5) & 34.635(2) & 17.6166(9) & - \\
                                 &         &        &        &        \\
$\bb b e^+ e^-            $ & 24.646(6) & 18.290(7) & 10.380(6)  & - \\
                                 &         &        &        &        \\
$\bb b \baru u            $ & 86.72(6)  & 62.43(7)  & 31.16(7)   & - \\
                                 &         &        &        &        \\
$\bb b \bard d            $ &109.8(1)   & 78.6(1)   & 38.4(1)    & - \\
                                 &         &        &        &        \\
\hline
\end{tabular}
\end{center}
\label{tab6}
\caption{Cross sections in fb for $e^+e^- \to \bb b \fb f$. The kinematical
cuts have been given explicitly in the previous section.}
\end{table}

\noindent
At this point we also present results for the cross section relative to
the process $e^+e^- \to \bb b \bb b$. Within our approximations, which
have already been described in the paper, we have adopted two different 
algorithms.

\begin{itemize}

\item[A1] In the first one all the $bb$ and $\bb b$ pairs are required to have
an invariant mass of at least $30\,$GeV to suppress the gluonic and
photonic components. 

\item[A2] In the second we have adopted the following
algorithm. Let $M_{ij}$ be the invariant mass of the final state $i-j$ pair 
($i,j = 1,4$), then we define $M_i(i=1,6)$ by
\vskip 0.5cm

\begin{eqnarray}
M_1 &=& M_{12}, \qquad M_2 = M_{34},   \nll
M_3 &=& M_{13}, \qquad M_4 = M_{24},   \nll
M_5 &=& M_{14}, \qquad M_6 = M_{23}.
\end{eqnarray}
\vskip 0.5cm

\noindent
Next we always require $M_i \geq 5\,$GeV and we further apply the rule

\begin{itemize}

\item[-] if $\{|M_1-\zm| < 25\,\gev, M_2 > 50\,\gev\}$.or.
           $\{|M_2-\zm| < 25\,\gev, M_1 > 50\,\gev\}$ then the event is accepted

\item[-] else if $\{|M_3-M_Z| < 25 \,\gev, M_3 > 50 \,\gev\}$.or.
                 $\{|M_4-M_Z| < 25 \,\gev, M_3 > 50 \,\gev\}$ then the event is
accepted

\item[-] else if $\{|M_5-M_Z| < 25 \,\gev, M_6 > 50 \,\gev\}$.or.
                 $\{|M_6-M_Z| < 25 \,\gev, M_5 > 50 \,\gev\}$ then the event is
accepted

\item[-] else the event is rejected.

\end{itemize}
\end{itemize}

\noindent
For a $\hm = 80,90\,$GeV the corresponding cross sections are shown in 
table $7$ where the first entry refers to A1 and the second to A2.

\begin{table}[hbtp]
\begin{center}
\begin{tabular}{|c|c|c|}
\hline
$\sqrt{s}\,$(GeV)/$\hm\,$(GeV)  & 80 & 90  \\
\hline
     &                  &               \\
160  &    1.973 & 1.486 \\
     &    5.050 & 4.078 \\
     &                  &               \\
165  &    2.713 & 1.730 \\
     &    6.706 & 4.959 \\
     &                  &               \\
170  &    6.004 & 2.122 \\
     &    12.181 & 6.006 \\
     &                  &               \\
175  &    22.592 & 2.924 \\
     &    38.665 & 7.858 \\
     &                  &               \\
180  &    30.995 & 6.342 \\
     &    54.133 & 38.555 \\
     &                  &               \\
\hline
\end{tabular}
\end{center}
\label{tab7}
\caption{Cross sections (fb) for $e^+e^- \to \bb b \bb b$. The first entry
correspond to the selection $M_{ij} \geq 30\,$GeV while the second one
refers to the algorithm described in the text.}
\end{table}

\noindent
There are several differential distributions which are of some relevance in the 
Higgs study. They provide informations useful for choosing cuts in the
Higgs searches. Among them we have selected:

\begin{itemize}

\item[-] The $M(\bb b)$ distribution for all channels but $\bb b \bb b$.
It is useful whenever the direct reconstruction of the invariant mass from
the jets in the process is viable.

\item[-] The $M(\fb f)$ distribution, in particular -- but not only -- the
$M(\bn\nu)$ one.

\item[-] The missing mass recoil. A knowledge of $\sqrt{s}$ and of the leptonic
final states is required
\vskip 0.5cm

\begin{equation}
M_{rec}^2 = s - 2\,\sqrt{s}\left(E_{l^+} + E_{l^-}\right) + M^2(l^+l^-)
\end{equation}
\vskip 0.5cm

\item[-] The visible energy in $\bb b\bn \nu$, or in general $E(\bb) +
E(b)$. The $b$-quark pairs from the Higgs decay have a sharp peak in the
energy distribution due to the small Higgs width.

\item[-] Angular distributions for the $b$-quark and/or the $\bb$-quark.
In particular the $\cos\theta(\bb b)$ distribution of the total $3$-momentum
${\vec p}_{\bb b}$. However, in general, the signal angular distributions are
very isotropic.

\end{itemize}

\noindent
Some care has been devoted in understanding qualitatively the effect of flavor
mis-identification. For instance we have considered the process
\vskip 0.5cm

\begin{eqnarray}
e^+e^- &\to & \mu^- \mu^+ b \bb,  \nll
e^+e^- &\to & \nu_{\mu} \bn_{\mu} b \bb.
\end{eqnarray}
\vskip 0.5cm

\noindent
In absence of flavor identification one has to consider more processes which
are subsequently weighted with some external probability. For instance
we have assumed the following weights~\cite{ob1}
\vskip 0.5cm

\begin{eqnarray}
e^+ e^- \to  \mu^-\mu^+ &b& \bb, \qquad   P_{bb}^2 = 0.4665,  \nll
                        &d& \bard, \qquad   P_{db}^2 = 0.0014,  \nll
                        &s& \bars, \qquad   P_{sb}^2 = 0.0029,  \nll
                        &u& \baru, \qquad   P_{ub}^2 = 0.0014,  \nll
                        &c& \barc, \qquad   P_{cb}^2 = 0.0818,  
\end{eqnarray}

\noindent
or

\begin{eqnarray}
e^+ e^- \to  \nu\bn  &b& \bb, \qquad   P_{bb}^2 = 0.4665,  \nll
                        &d& \bard, \qquad   P_{db}^2 = 0.0014,  \nll
                        &s& \bars, \qquad   P_{sb}^2 = 0.0029,  \nll
                        &u& \baru, \qquad   P_{ub}^2 = 0.0014,  \nll
                        &c& \barc, \qquad   P_{cb}^2 = 0.0818.
\end{eqnarray}
\vskip 0.5cm

\noindent
The above probabilities are just for an indication of the general idea which
we have illustrated.
The gross features of the differential distributions can be understood from the
structure of the sharp peaks around different values of different invariant 
masses. Typically the narrow width of the Higgs boson will be reflected
by an unmistakable peak at $M(\bb b) = \hm$. It goes without saying that
our predictions are at the partonic level and that the experimental resolution
(energy of the $b$ quarks, angles etc.) has not been included. The latter when
properly included will inevitably reduce this rather spectacular peak.

From the full set of diagrams contributing to different channels we have
peaks around zero values of some invariant mass due to the sub-processes
$\gamma^*(g) \to \fb f$ which are usually eliminated by cutting the low 
values of that variable. We also have peaks in invariant masses around
$\zm$ due to $Z^* \to \fb f$ which becomes dominant whenever the energy
becomes larger and larger and the $ZZ$ background component increases.

There are also peaks around the beam axis due to $t$-channel diagrams
as the already mentioned multi-peripheral contributions in $e^+e^- \to
\bb b e^+e^-$. Finally we do not attribute any particular relevance to
a separation of the Higgs signal into Higgsstrahlung or fusion components,
they are both present in any complete calculation.

For all processes $e^+e^- \to \bb b \fb f, f \not= b$ we have shown the
$M(\bb b)$ distribution for $\sqrt{s} = 175\,$GeV, $\hm = 80\,$GeV while
for $\sqrt{s} = 190\,$GeV we have considered three values
of the Higgs boson mass, $\hm = 80,90$ and $100\,$GeV. They are given
in figure $4$ through figure $7$.

The relative importance of the ratio signal/background from the point of view 
of the $M(\bb b)$ distribution is given for a Higgs of $80\,$GeV and
$\sqrt{s} = 190\,$GeV, in figure $8$.

Flavor mis-identification has been analyzed in figure $9$ 
and $10$ where we have considered $e^+e^- \to \mu^+ \mu^- (\bn_{\mu}
\nu_{\mu}) \barq q$. They clearly show the reduction of the $\bb b$
signal and the contamination around $M(\barq q) = \zm$ from the $\barc c$
mis-identification, the rest remaining negligible.

The ratio signal/background is better illustrated in terms of other
differential distributions, typically $M_{miss}$ or $E_{\bb+b}$.
Usually the $M_{miss}$ distribution is used for $e^+e^- \to \bb b l^+ l^-$
only but we have also shown its behavior for $e^+e^- \to \bb b \barq q$
in a situation where exact flavor identification is assumed and
$M(\bb b) \geq 50\,$GeV, $\mid M(\barq q) - \zm\mid \leq 25\,$GeV.
As it appears the $b$-quark pairs from the Higgs decay have a sharp peak in the
energy distribution due to the small Higgs width.

For all processes $e^+e^- \to \bb b \fb f, f \not= b,\nu$ we have shown the
$M_{miss}$ distribution for $\sqrt{s} = 175\,$GeV, $\hm = 80\,$GeV and for
$\sqrt{s} =190\,$GeV, $\hm = 80,90$ and $100\,$GeV. They are given
in figure $11$ through figure $14$. Similarly for all $f \not= b$
-- thus assuming flavor identification -- we have shown the energy distribution
in figure $15$ through figure $18$.

The signal/background ratio is given in terms of $M_{miss}$ and of the 
$b$-quark energies in figure $19$ and in figure $20$.
Finally we have reported in figure $21$ the differential distribution 
in $\cos\theta_{_H}$, where $\theta_{_H}$ is the angle formed by 
${\vec p}(\bb) + {\vec p}(b)$ with the beam direction. As anticipated one 
can see that, in general, the signal angular distributions are very isotropic 
for all channels.

A final comment has to be devoted to the effective inclusion of the $b$-quark 
mass. Using the approximate formulation of eq.~\ref{emass} we have estimated
this effect by considering the $M(\bb b)$ distribution in $e^+e^- \to
\bb b \bn_e \nu_e$ at $\sqrt{s} = 175\,$GeV and $\hm = 80\,$GeV.
The deviation of the ratio (effective mass)/(massless) from one for this 
quantity is of the order of $0.1\%$ around $M(\bb b) = \hm$ and hardly
noticeable away from it. As for the cross section we have taken the
process $e^+e- \to \bb b \mu^+ \mu^-$ and compared the effective mass
treatment with the massless case in table $8$.
With respect to the results presented by CompHEP and WPHACT in ref.~\cite{yr2},
which take $m_b$ exactly into account, it should be said that they use
$m_b = m_b($pole$) = 4.7\,$GeV in the matrix elements while here
$m_b(80\div 90\,\gev) \approx 2.7\,$GeV. The choice of ref.~\cite{yr2}
is the standard one which was agreed upon and used during the LEP~2 workshop.
Of course no general statement can be made here about the full $m_b$-mass 
dependence and to a large extent the size of the effect is deeply related to 
the particular channel under examination and on the chosen set of kinematical 
cuts.
Therefore the results shown in table $8$ are only valid for the 
particular setup under consideration and should not be confused with a process
independent statement. To the best of our knowledge the general answer is under
examination by the authors of WPHACT~\cite{pc}.

\begin{table}[hbtp]
\begin{center}
\begin{tabular}{|c|c|c|}
\hline
$\sqrt{s}\,$(GeV)/$\hm\,$(GeV)  & 80 & 90  \\
\hline
    &            &             \\
175 &  9.243(1)  &  1.1291(8)  \\
    &  9.232(1)  &  1.1288(8)  \\
    &  -1.2      &  -0.3       \\
    &            &             \\
190 &  22.339(3) &  16.213(1)  \\
    &  22.320(3) &  16.204(1)  \\
    &  -0.9      &  -0.6       \\
    &            &             \\
\hline
\end{tabular}
\end{center}
\label{tab8}
\caption{Cross sections (fb) for $e^+e^- \to \bb b \mu^+ \mu^-$. The first 
entry corresponds to massless $b$-quarks while the second one
refers to the inclusion of $m_b$ following eq.~\ref{emass}. The last entry 
gives the relative deviation in permill.}
\end{table}
 
\section{Conclusions}

The combined efforts of many different groups, from Europe to Japan, has 
shown during the last
years that the agreement between the dedicated Higgs codes is systematically 
better than $1\%$. Here we are considering those theoretical predictions
having control over the full set of diagrams which contribute to a given 
channel. Our point of view is very simple in this respect, it is
important to stress that the impact of the discrepancies between dedicated
or general purpose calculations is minimal on the discovery potential of 
LEP~2. 

However there is another important question to be answered, the extraction of 
Higgs properties from the experimental data. Here the perspectives change
since we move from discovery physics to precision physics and a full
control on the exact matrix elements is required to eliminate any source
of large discrepancies.
Once this is done we are left with the effect of additional $\%$ level 
uncertainties which are of the same order of those coming from higher order 
corrections. To understand the remaining small discrepancies will require 
further work.

Although a theoretical error by itself is not a well defined quantity
we have attempted some very primitive analysis of the problem, trying
to understand the interplay between variations in the results among different
renormalization schemes and the choice of the cuts which is more suitable
to minimize the corresponding effects.

Our main motivation in this paper has been twofold. First we start from the
observation that the presently available ensemble of experimental data is
not inconsistent with a (minimal standard model) Higgs boson within the range
of LEP~2 energies. Secondly the guidance of the published
comparisons for complete calculations on Higgs physics at LEP~2 shows that the 
achieved technical precision and the level of agreement are more than enough 
to motivate an extension of the previous work to include a larger number of 
channels in $e^+e^- \to \bb b \fb f$. Additional work is still needed in order
to have full control over the background, expecially for the $\bb b \bn \nu$
channel.

From a general point of view there is a need to move beyond the mere 
calculation of the total cross sections, different kinds of differential 
distributions must be made available to access the widest information for 
choosing cuts, for discussing the physics of a Higgs boson with a mass around 
$90\,$GeV where the Higgs and the $Z$ signal become degenerate and for a 
comprehensive analysis of the various background components. In this respect 
our attempt has been to enlarge several analyses already published in the 
literature.

In this context we have considered all channels $e^+e^- \to \bb b \fb f$
with $m_f = 0$ everywhere and $m_b$ non zero only in the Yukawa coupling.
Starting from the matrix elements at the parton level we have computed
cross sections and differential distributions for various c.m. energies
and (minimal standard model) Higgs boson masses. The logical steps to be
followed in this field are 

\begin{itemize}

\item[{\bf a}] to include the matrix elements for the full $4$-fermion process 
$e^+e^- \to \bb b \fb f$ beyond the factorization approximation.
For four quarks in the final state QCD processes must be added.

\item[{\bf b}] The mass of the $b$-quark should in principle be kept in the 
matrix elements, developing a finite interference between signal and background.
In our opinion the relatively small value of $m_b(\hm)$ is enough to
justify the massless approximation.

\item[{\bf c}] A description of the basic $4$-fermion process beyond the 
minimal standard model, such as SUSY models~\cite{susy}, should be made 
available.

\item[{\bf d}] After a description of the relevant differential distributions at
the parton level the unweigthed events with the $4$-momenta of all final state 
particles should be provided in order to process the events by applying 
analysis cuts.

\end{itemize}

\noindent
In our analysis we have fulfilled both - a - and - d - of the previous list.
We have also indicated some approximations to be used for a quick estimate of
the $b$-quark mass effect and discussed their validity. Although
unweigthed events have actually been generated no effort has been made during 
this work for a proper interface with the hadronization packages.

The outcome of our work is illustrated by several tables and figures which 
show the feasibility of the project. Among several technical aspects which
we consider as extremely relevant for any detailed discussion of the
standard Higgs boson properties the main conclusion reiterates the message that
if the Higgs mass is above $80\,GeV$ then the cross section is too small at 
$\sqrt{s} = 175\,$GeV to allow for a Higgs discovery, thus the $\sqrt{s} = 
190\,$GeV phase of the collider -- or even a higher one -- will be needed.

\section*{Acknowledgements}

I acknowledge many important discussions with Alessandro Ballestrero and
his active contribute in exchanging results prior to their publication.
His help was essential in creating the event generator branch of WTO and
in a better understanding of some of the sources of theoretical error.
Several important discussions with the members of the Higgs Physics
and of the Event Generators for Discovery Physics working groups at the recent
LEP~2 workshop are acknowledged. 
I sincerely acknowledge the active contribute of D.~Bardin, M.~Dubinin,
O.~Nicrosini and R.~Pittau in the comparison phase with their codes where we 
have been able to reach agreement with very high precision results on a sample
of processes.  

\section{Figure Captions}

\begin{itemize}

\item[Fig. 1] Cross sections (pb) for $e^+e^- \to \bb b \fb f$ where $f = 
\mu,\nu_{\mu} (HZ$-component), $f = \nu_e (HZ + WW$-components) and 
$f = e (HZ + ZZ$-components). Here $\hm = 80\,$GeV.
\vskip 0.5cm

\item[Fig. 2] Cross sections (pb) for $e^+e^- \to \bb b \bn n$ for
$\hm = 80,90\,$GeV.
\vskip 0.5cm

\item[Fig. 3] Cross sections (pb) for $e^+e^- \to \bb b \fb f$ for
$\hm = 80,90,100\,$GeV and $\hm = \infty$.
\vskip 0.5cm

\item[Fig. 4] The $M(\bb b)$ distribution for all processes $e^+e^- \to \bb b 
\fb f, f \not= b$ at $\sqrt{s} = 175\,$GeV and $\hm = 80\,$GeV.  
\vskip 0.5cm

\item[Fig. 5] The same as in Fig. 4 but with $\sqrt{s} = 190\,$GeV and 
$\hm = 80\,$GeV.
\vskip 0.5cm

\item[Fig. 6] The same as in Fig. 4 but with $\sqrt{s} = 190\,$GeV and 
$\hm = 90\,$GeV.
\vskip 0.5cm

\item[Fig. 7] The same as in Fig. 4 but with $\sqrt{s} = 190\,$GeV and 
$\hm = 100\,$GeV.
\vskip 0.5cm

\item[Fig. 8] The $M(\bb b)$ distribution for all processes $e^+e^- \to \bb b 
\fb f, f \not= b$ at $\sqrt{s} = 190\,$GeV and $\hm = 80\,$GeV.  
The Higgs boson signal and its background are compared.
\vskip 0.5cm

\item[Fig. 9] Flavor mis-identification is considered in the process
$e^+e^- \to \mu^+ \mu^- \barq q$ for $\sqrt{s} = 175\,$GeV and $\hm = 80\,$GeV.
\vskip 0.5cm

\item[Fig. 10] The same as in Fig. 9 but for $e^+e^- \to \bn_{\mu}
\nu_{\mu} \barq q$.
\vskip 0.5cm

\item[Fig. 11] The $M_{miss}$ distribution for all processes $e^+e^- \to \bb b 
\fb f, f \not= b,\nu$ at $\sqrt{s} = 175\,$GeV and $\hm = 80\,$GeV.  
\vskip 0.5cm

\item[Fig. 12] The same as in Fig. 11 but with $\sqrt{s} = 190\,$GeV and 
$\hm = 80\,$GeV.
\vskip 0.5cm

\item[Fig. 13] The same as in Fig. 11 but with $\sqrt{s} = 190\,$GeV and 
$\hm = 90\,$GeV.
\vskip 0.5cm

\item[Fig. 14] The same as in Fig. 11 but with $\sqrt{s} = 190\,$GeV and 
$\hm = 100\,$GeV.
\vskip 0.5cm

\item[Fig. 15] The $E_{\bb+b}$ distribution for all processes $e^+e^- \to \bb b 
\fb f, f \not= b$ at $\sqrt{s} = 175\,$GeV and $\hm = 80\,$GeV.  
\vskip 0.5cm

\item[Fig. 16] The same as in Fig. 15 but with $\sqrt{s} = 190\,$GeV and 
$\hm = 80\,$GeV.
\vskip 0.5cm

\item[Fig. 17] The same as in Fig. 15 but with $\sqrt{s} = 190\,$GeV and 
$\hm = 90\,$GeV.
\vskip 0.5cm

\item[Fig. 18] The same as in Fig. 15 but with $\sqrt{s} = 190\,$GeV and 
$\hm = 100\,$GeV.
\vskip 0.5cm

\item[Fig. 19] The $M_{miss}$ distribution for all processes $e^+e^- \to \bb b 
\fb f, f \not= b,\nu$ at $\sqrt{s} = 190\,$GeV and $\hm = 80\,$GeV.  
The Higgs boson signal and its background are compared.
\vskip 0.5cm

\item[Fig. 20] The $E_{\bb+b}$ distribution for all processes $e^+e^- \to \bb b 
\fb f, f \not= b$ at $\sqrt{s} = 190\,$GeV and $\hm = 80\,$GeV.  
The Higgs boson signal and its background are compared.
\vskip 0.5cm

\item[Fig. 21] The differential distribution in $\cos\theta_{_H}$ for all 
processes $e^+e^- \to \bb b \fb f, f \not= b$ at $\sqrt{s} = 175\,$GeV and 
$\hm = 80\,$GeV. Here $\theta_{_H}$ is the angle formed by ${\vec p}_{\bb} + 
{\vec p}_b$ with the beam direction. The Higgs boson signal and its background 
are compared.

\end{itemize}

\newpage


\newpage

\begin{figure}[htbp]
\vspace{0.1cm}
\centerline{
\epsfig{figure=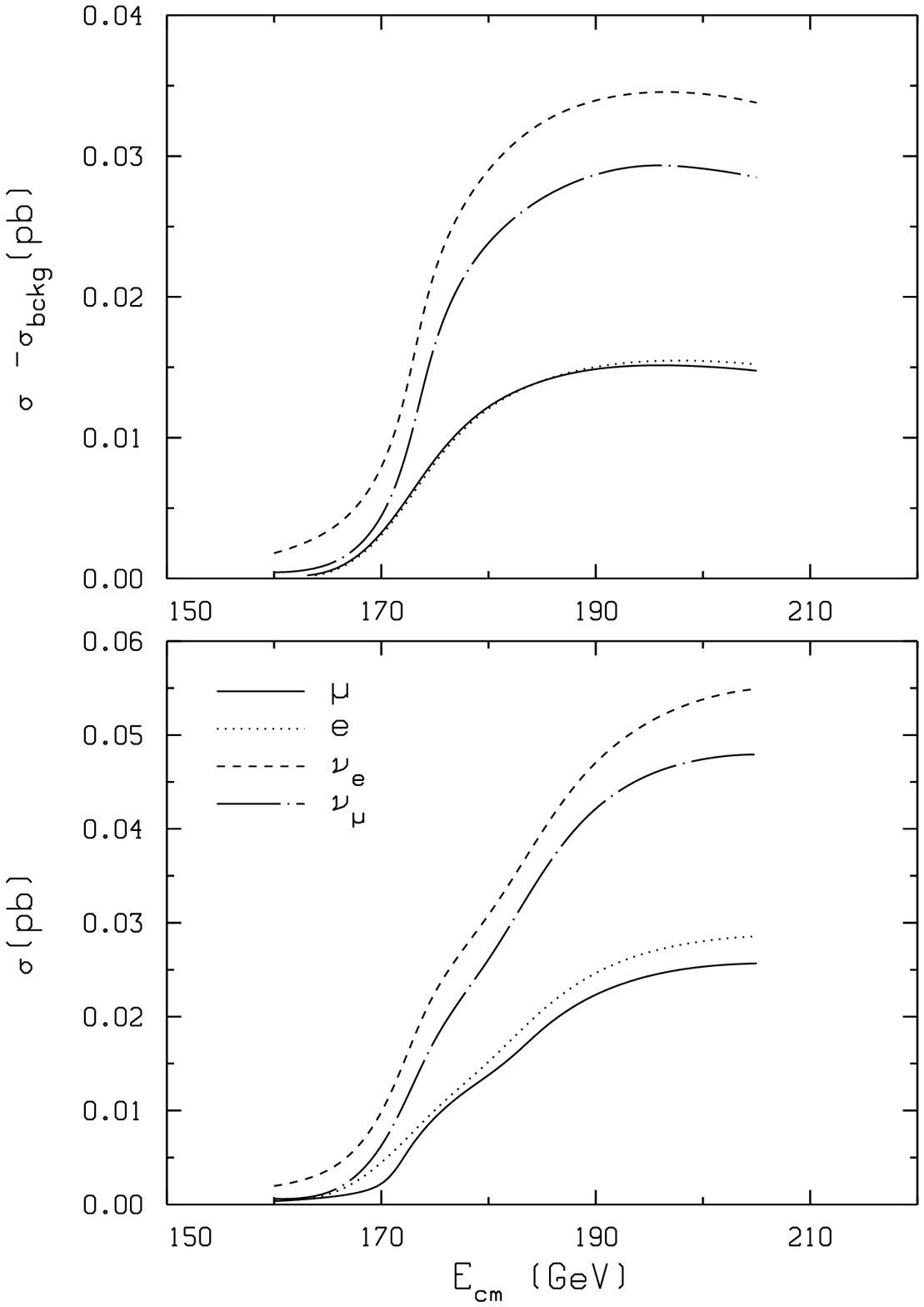,height=16cm,angle=0}
}
\vspace{0.1cm}
\begin{center}
{\bf Fig. 1}
\end{center}
\label{fga}
\end{figure}

\newpage

\begin{figure}[htbp]
\vspace{0.1cm}
\centerline{
\epsfig{figure=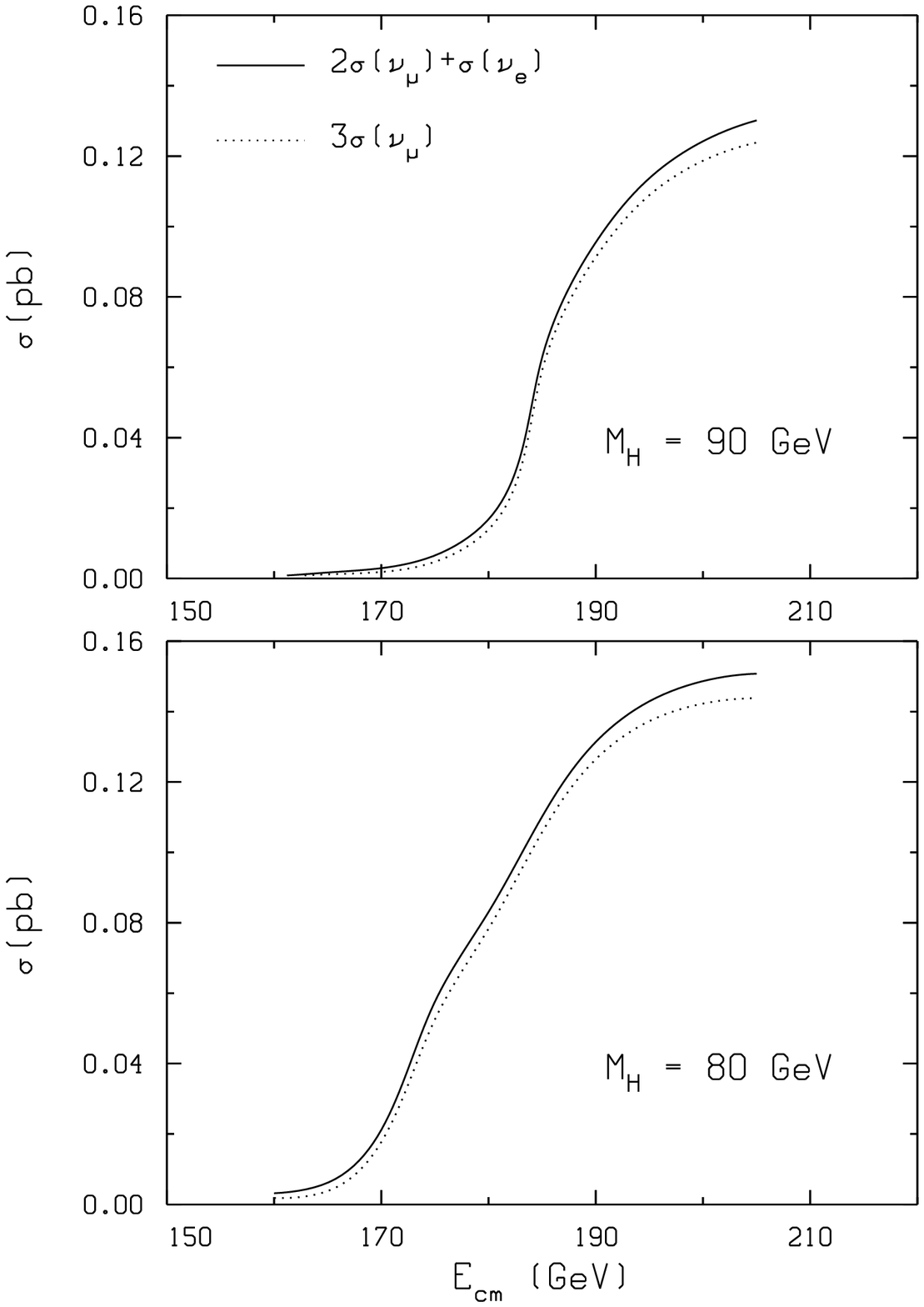,height=16cm,angle=0}
}
\vspace{0.1cm}
\begin{center}
{\bf Fig. 2}
\end{center}
\label{fgb}
\end{figure}

\newpage

\begin{figure}[htbp]
\vspace{0.1cm}
\centerline{
\epsfig{figure=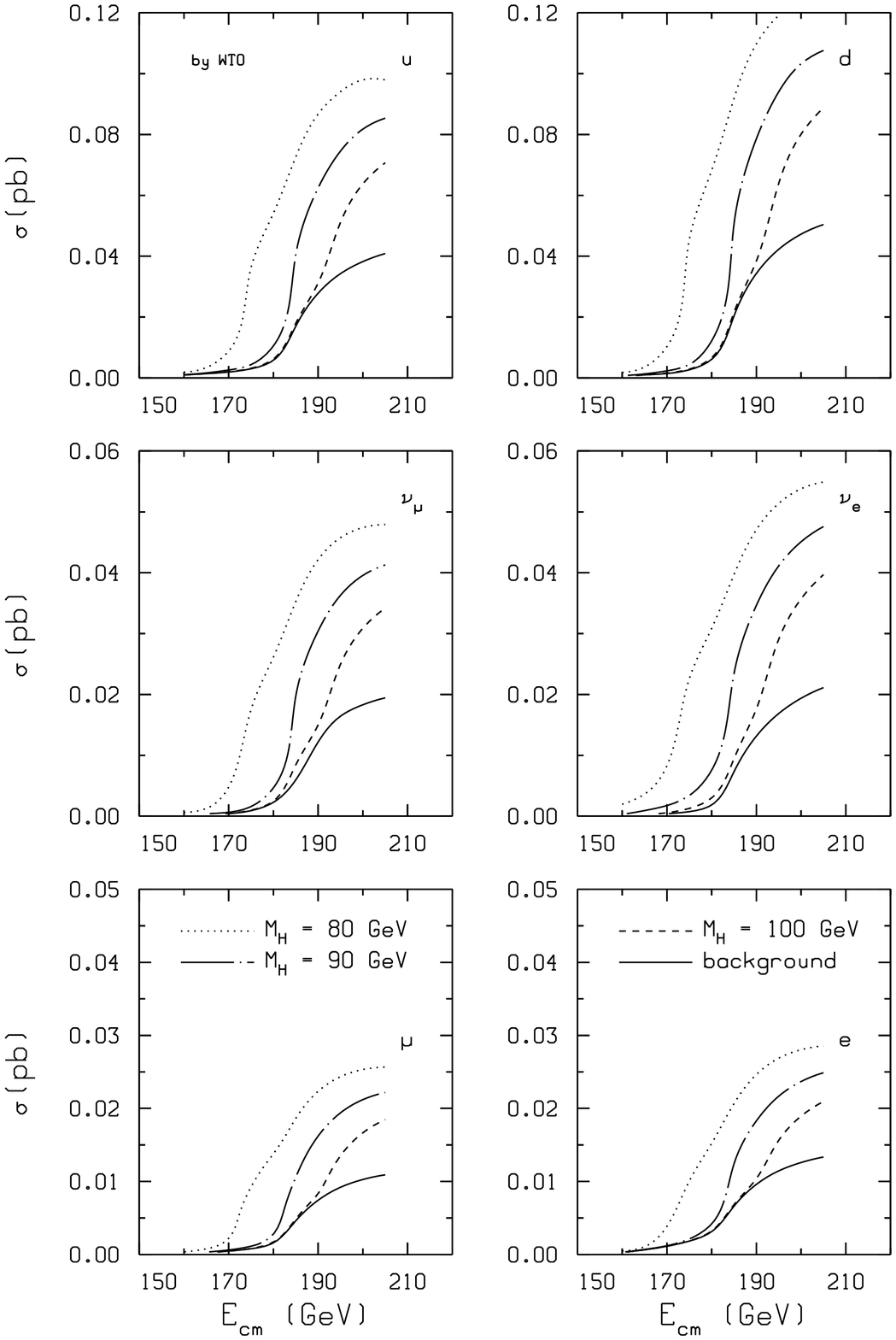,height=16cm,angle=0}
}
\vspace{0.1cm}
\begin{center}
{\bf Fig. 3}
\end{center}
\label{fgc}
\end{figure}

\newpage

\begin{figure}[htbp]
\vspace{0.1cm}
\centerline{
\epsfig{figure=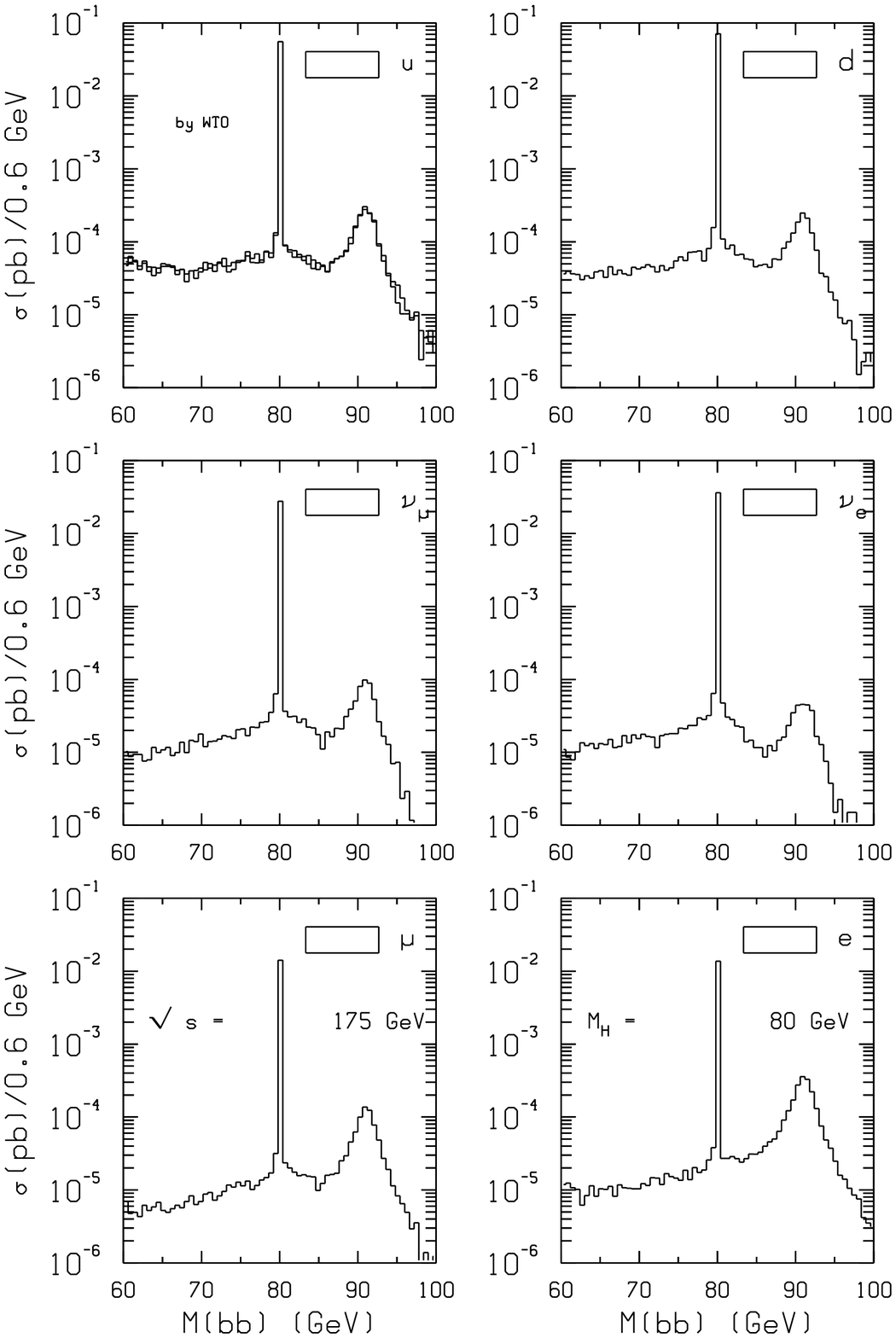,height=16cm,angle=0}
}
\vspace{0.1cm}
\begin{center}
{\bf Fig. 4}
\end{center}
\label{fgd}
\end{figure}

\newpage

\begin{figure}[htbp]
\vspace{0.1cm}
\centerline{
\epsfig{figure=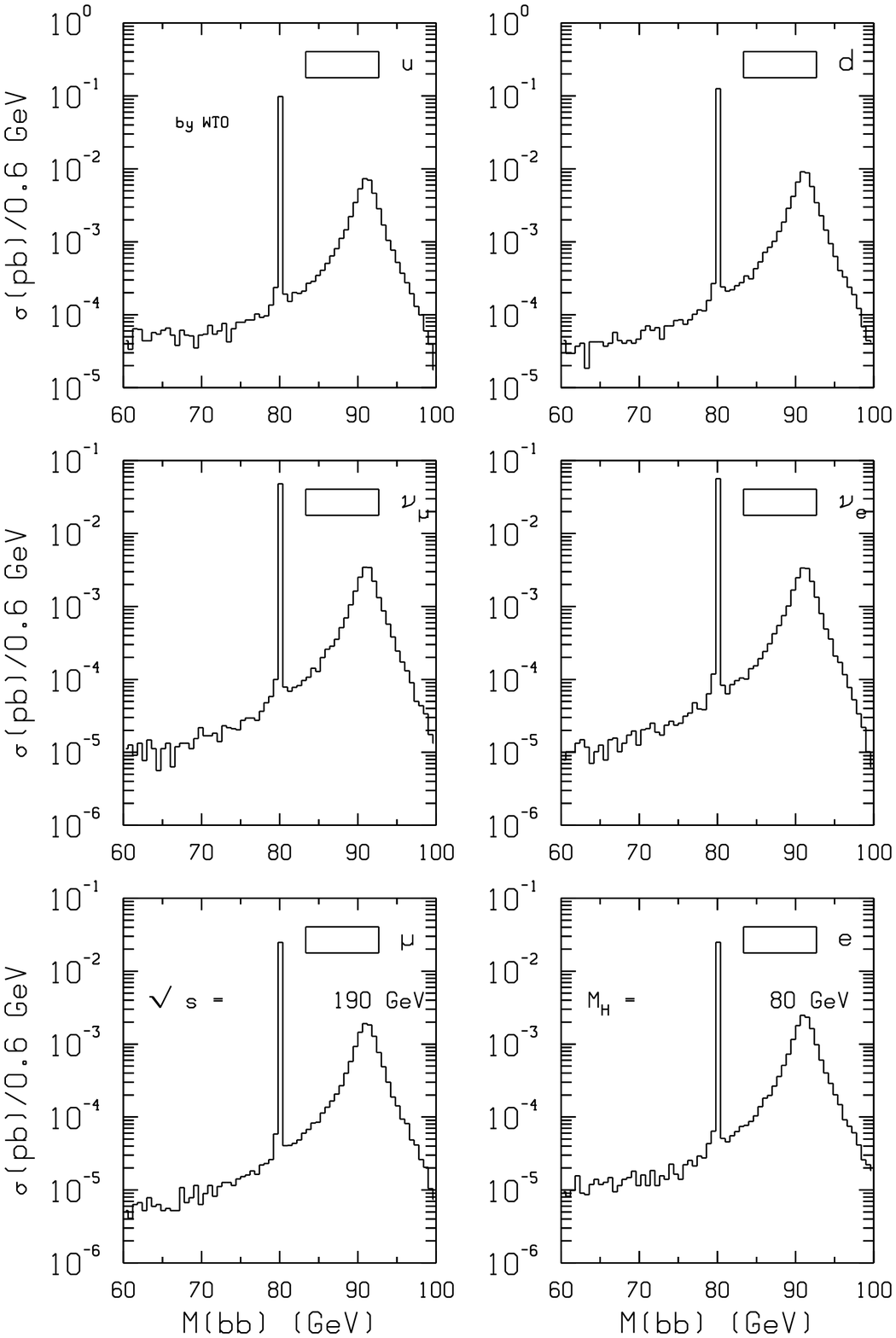,height=16cm,angle=0}
}
\vspace{0.1cm}
\begin{center}
{\bf Fig. 5}
\end{center}
\label{fge}
\end{figure}

\newpage

\begin{figure}[htbp]
\vspace{0.1cm}
\centerline{
\epsfig{figure=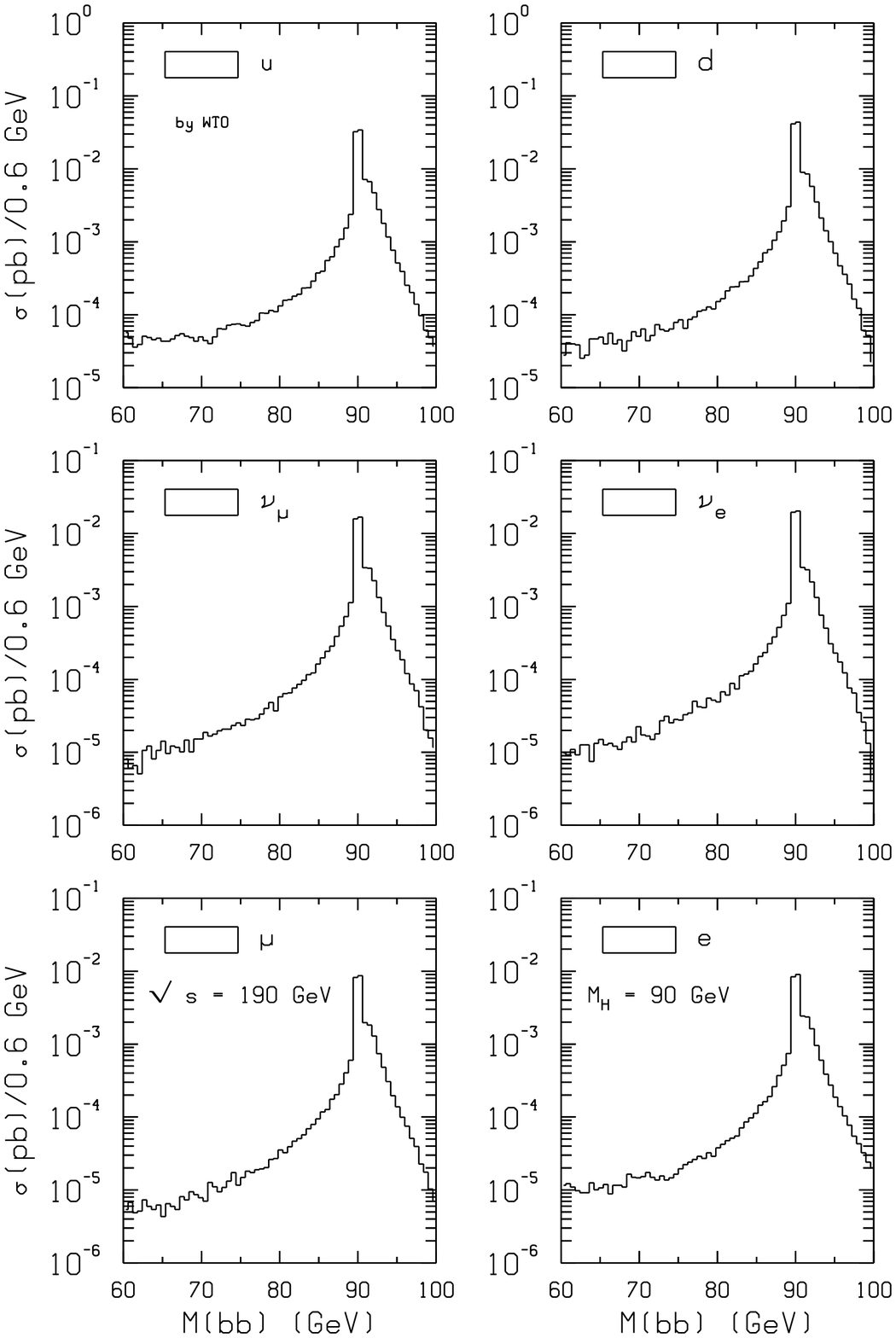,height=16cm,angle=0}
}
\vspace{0.1cm}
\begin{center}
{\bf Fig. 6}
\end{center}
\label{fgf}
\end{figure}

\newpage

\begin{figure}[htbp]
\vspace{0.1cm}
\centerline{
\epsfig{figure=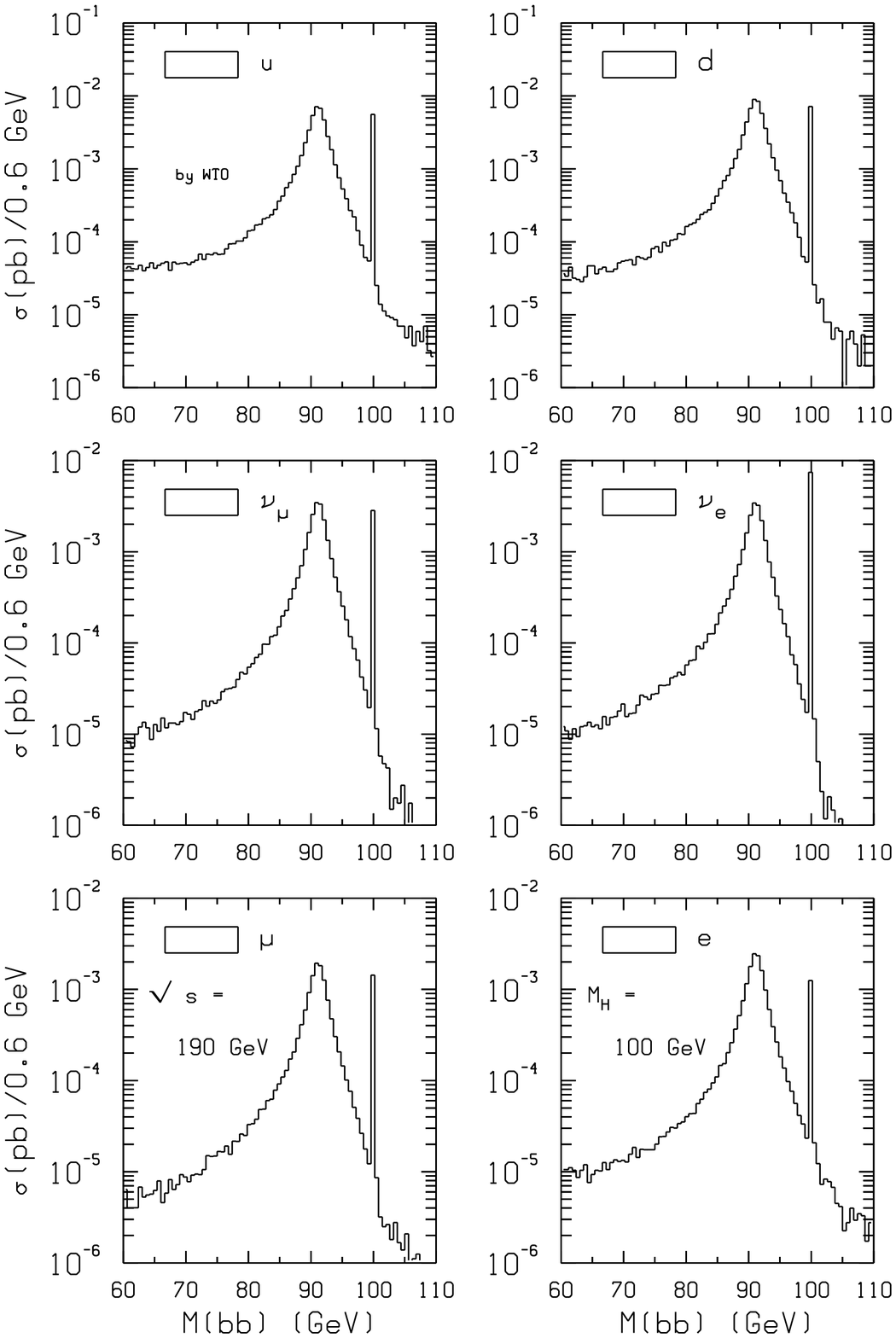,height=16cm,angle=0}
}
\vspace{0.1cm}
\begin{center}
{\bf Fig. 7}
\end{center}
\label{fgg}
\end{figure}

\newpage

\begin{figure}[htbp]
\vspace{0.1cm}
\centerline{
\epsfig{figure=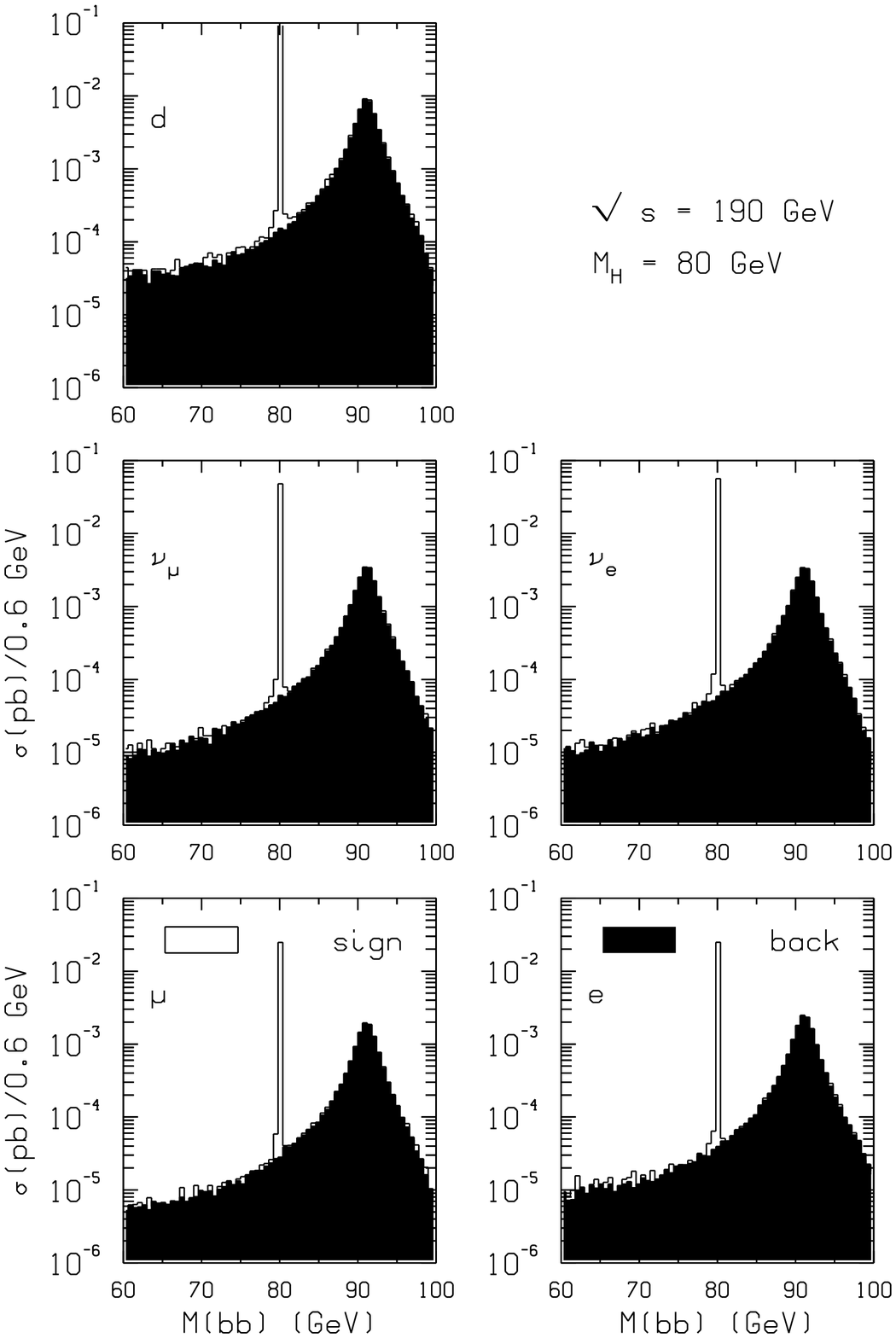,height=16cm,angle=0}
}
\vspace{0.1cm}
\begin{center}
{\bf Fig. 8}
\end{center}
\label{fgh}
\end{figure}

\newpage

\begin{figure}[htbp]
\vspace{0.1cm}
\centerline{
\epsfig{figure=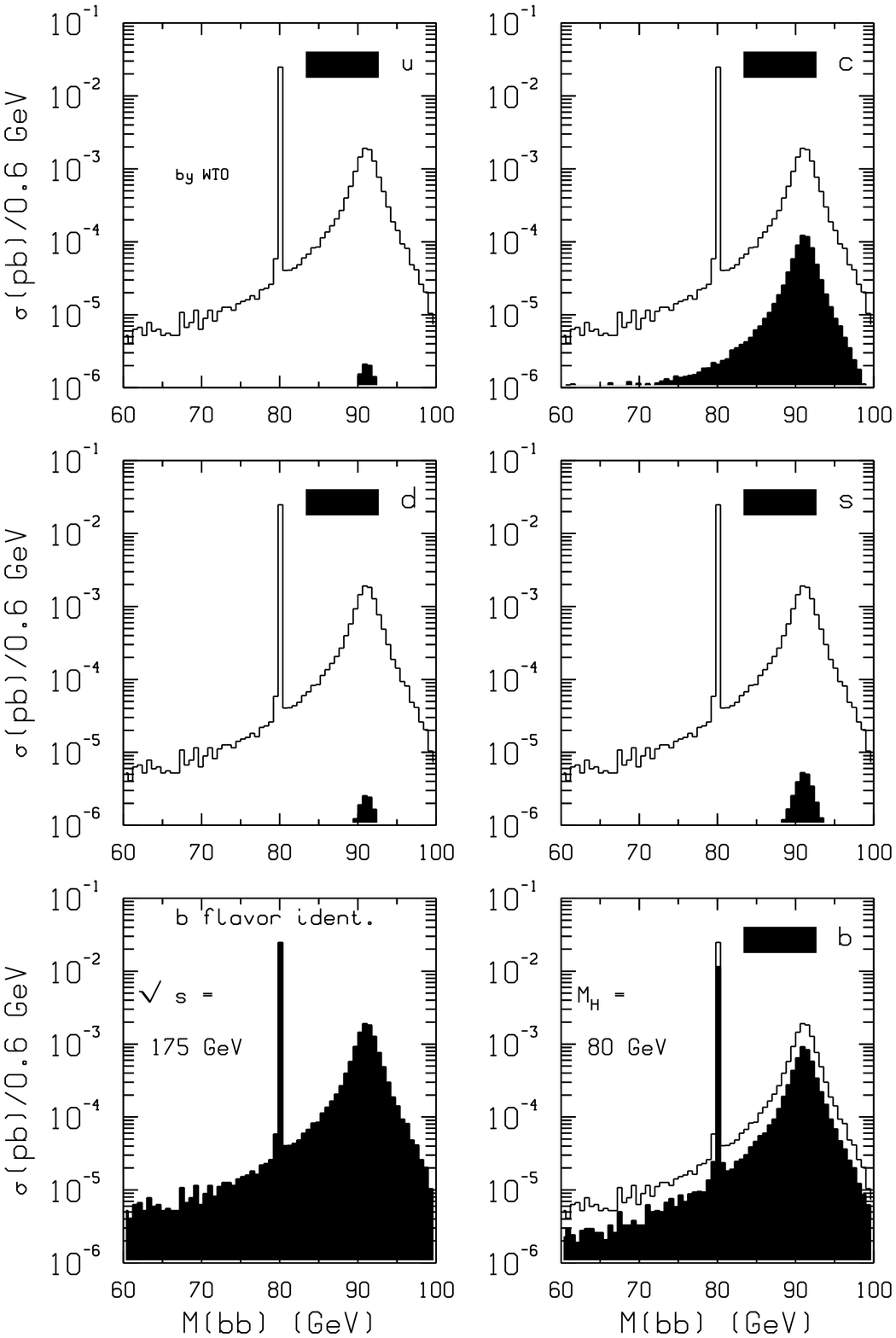,height=16cm,angle=0}
}
\vspace{0.1cm}
\begin{center}
{\bf Fig. 9}
\end{center}
\label{fgi}
\end{figure}

\newpage

\begin{figure}[htbp]
\vspace{0.1cm}
\centerline{
\epsfig{figure=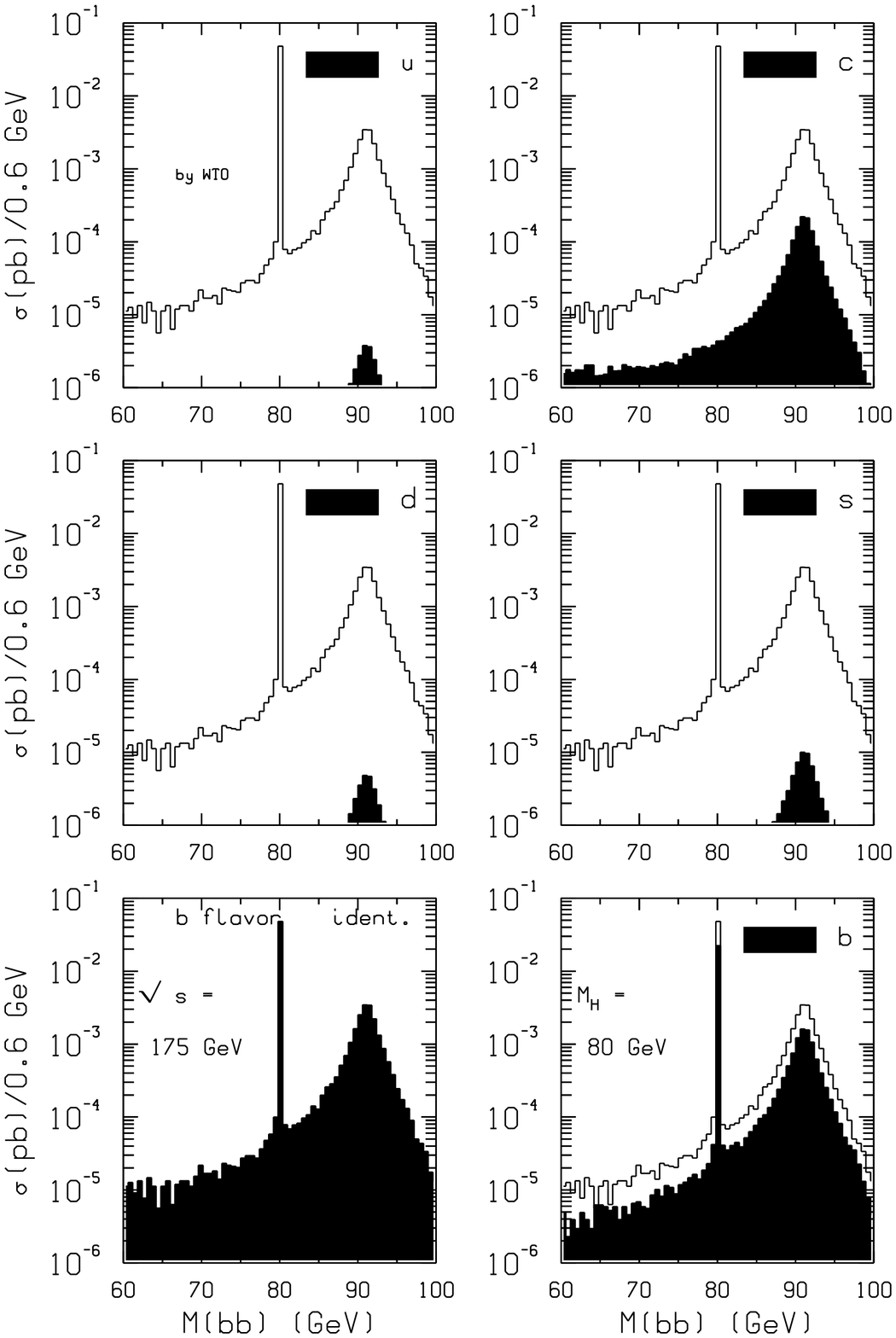,height=16cm,angle=0}
}
\vspace{0.1cm}
\begin{center}
{\bf Fig. 10}
\end{center}
\label{fgl}
\end{figure}

\newpage

\begin{figure}[htbp]
\vspace{0.1cm}
\centerline{
\epsfig{figure=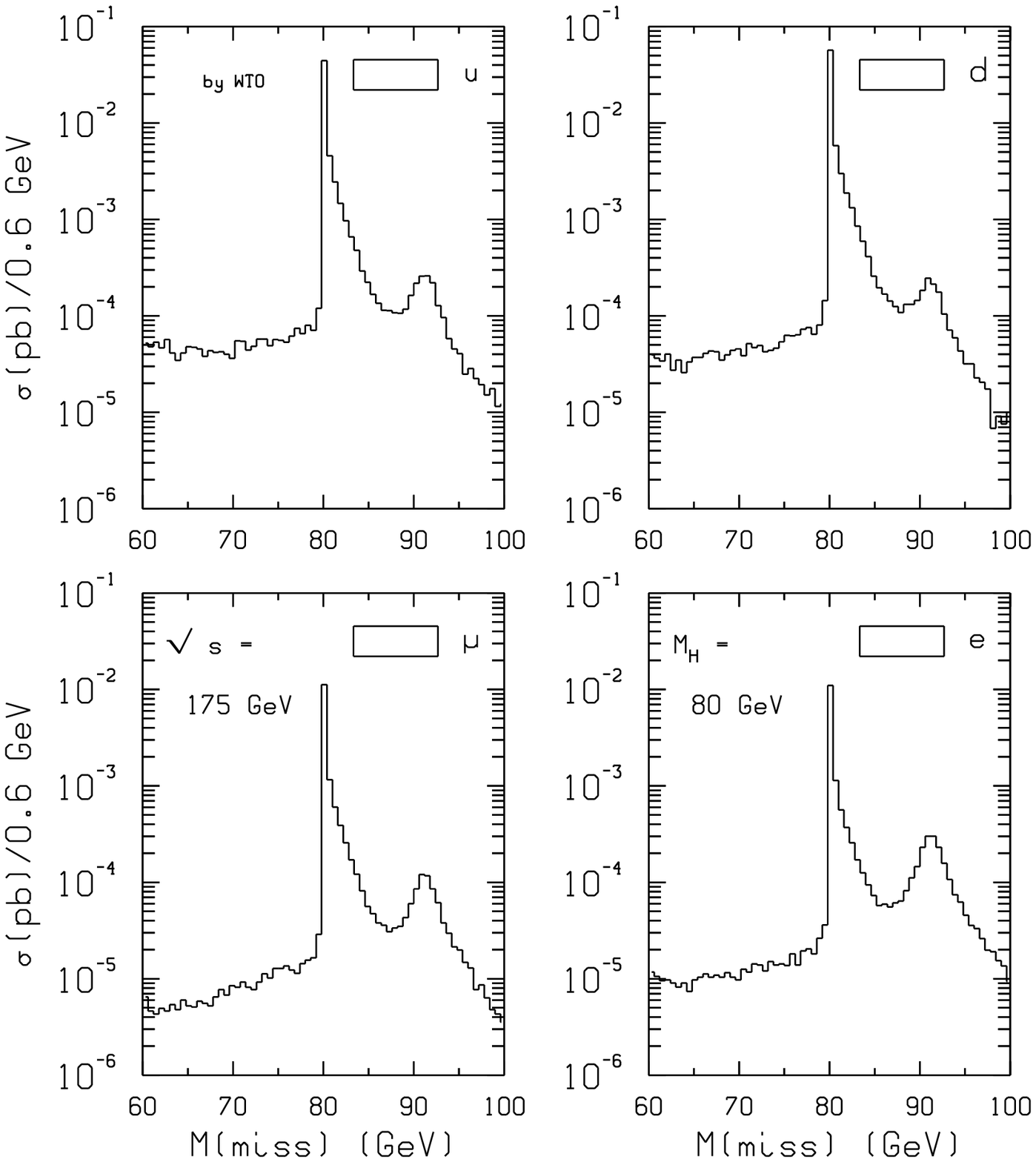,height=16cm,angle=0}
}
\vspace{0.1cm}
\begin{center}
{\bf Fig. 11}
\end{center}
\label{fgm}
\end{figure}

\newpage

\begin{figure}[htbp]
\vspace{0.1cm}
\centerline{
\epsfig{figure=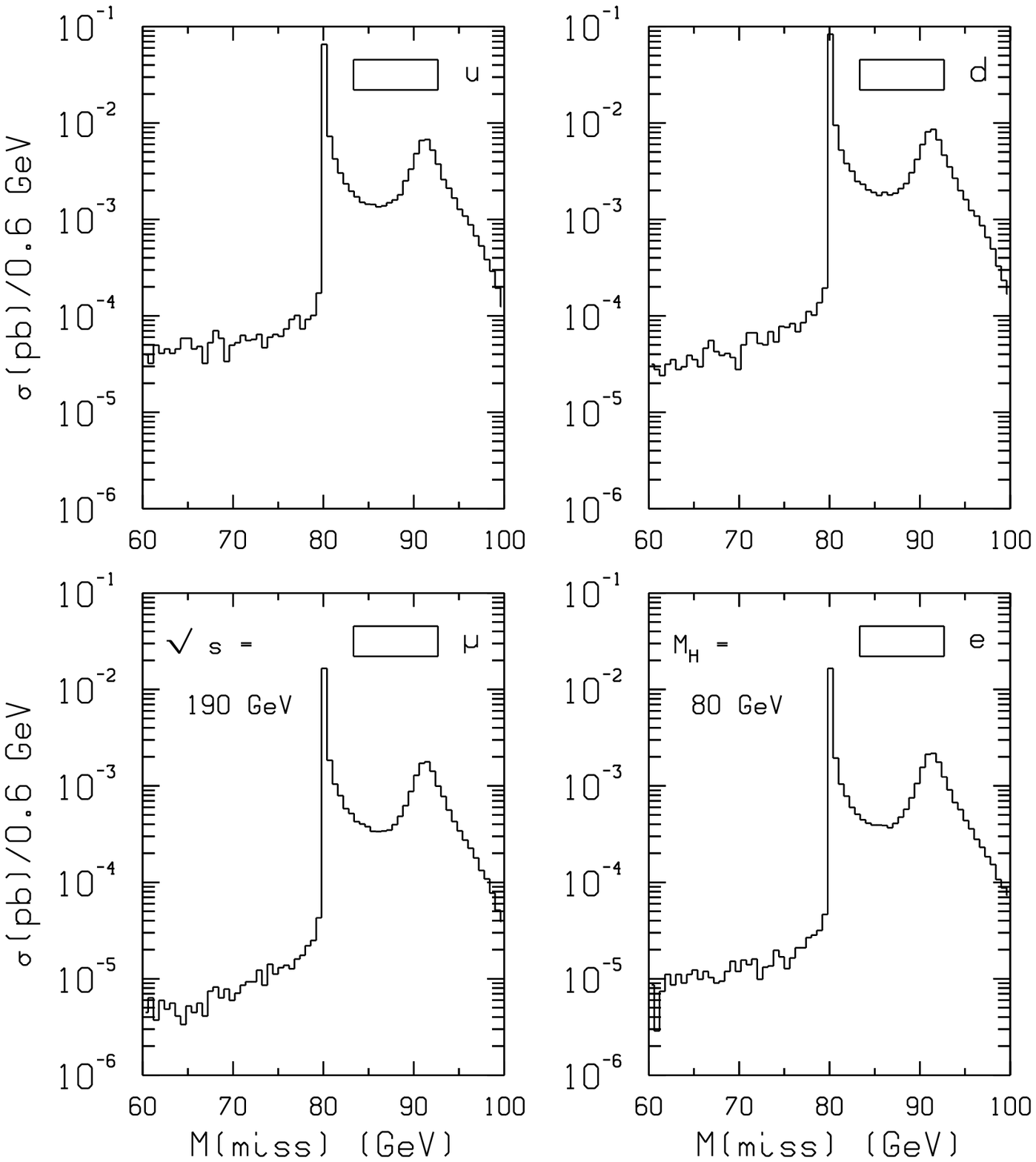,height=16cm,angle=0}
}
\vspace{0.1cm}
\begin{center}
{\bf Fig. 12}
\end{center}
\label{fgn}
\end{figure}

\newpage

\begin{figure}[htbp]
\vspace{0.1cm}
\centerline{
\epsfig{figure=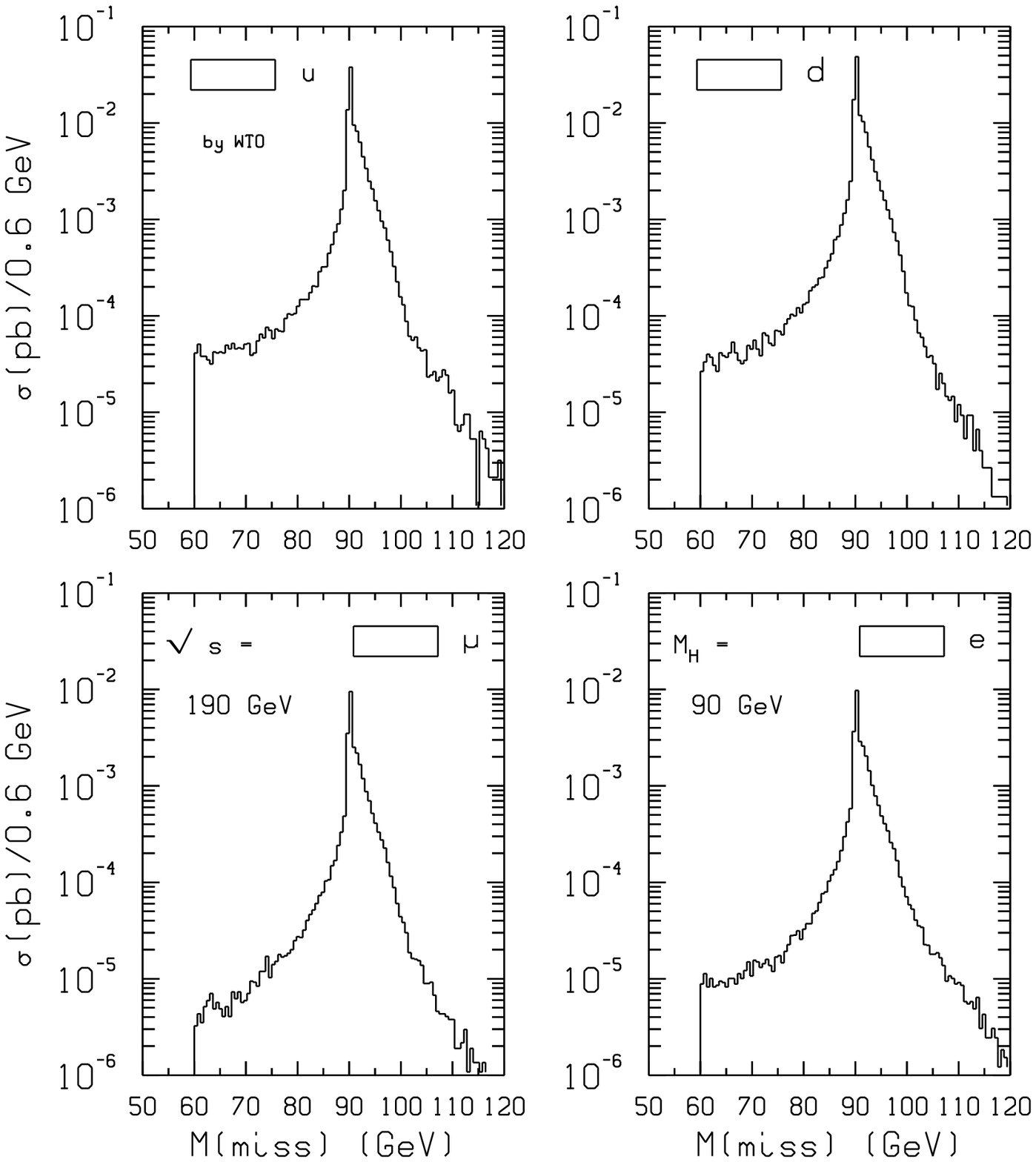,height=16cm,angle=0}
}
\vspace{0.1cm}
\begin{center}
{\bf Fig. 13}
\end{center}
\label{fgo}
\end{figure}

\newpage

\begin{figure}[htbp]
\vspace{0.1cm}
\centerline{
\epsfig{figure=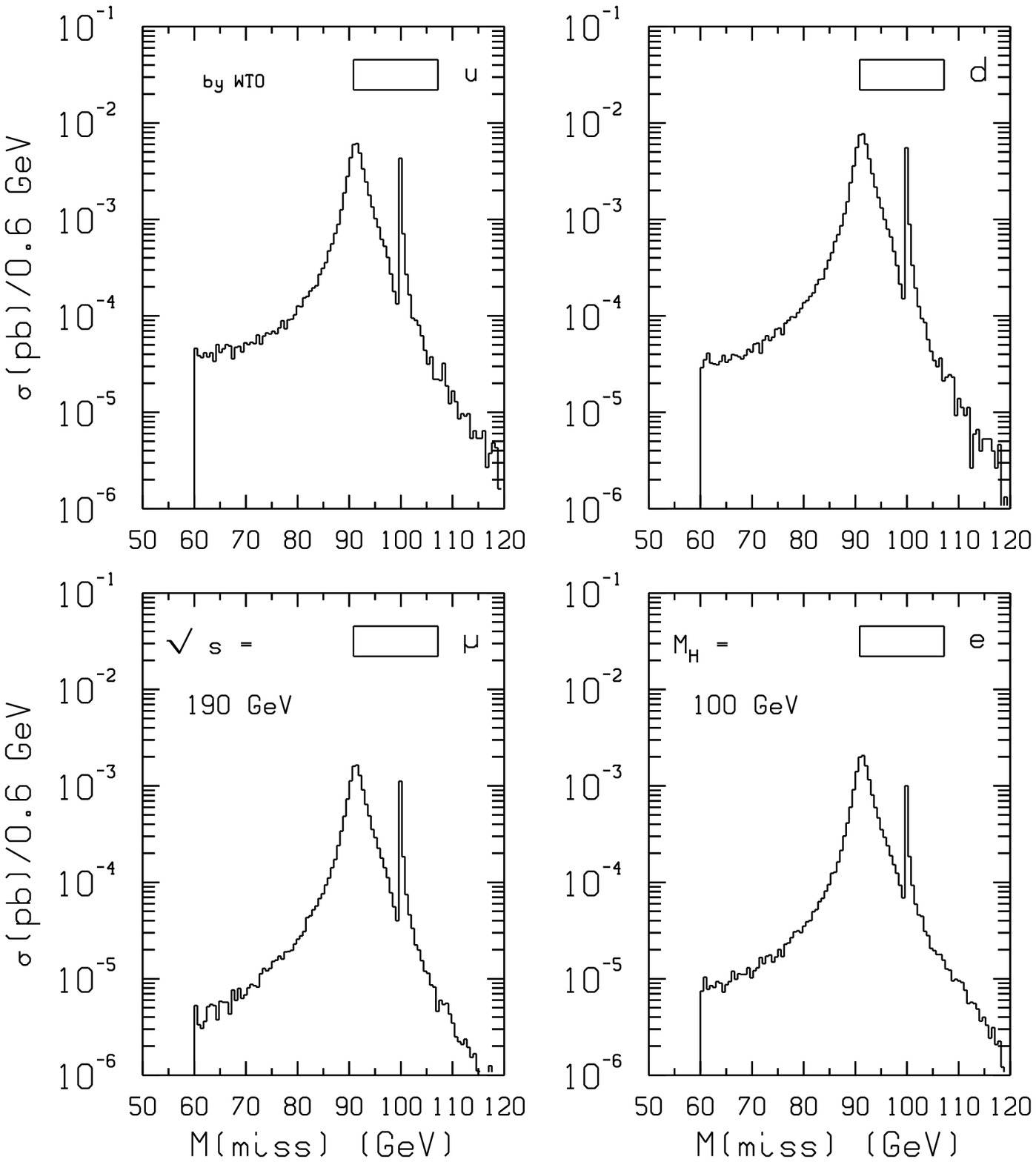,height=16cm,angle=0}
}
\vspace{0.1cm}
\begin{center}
{\bf Fig. 14}
\end{center}
\label{fgp}
\end{figure}

\newpage

\begin{figure}[htbp]
\vspace{0.1cm}
\centerline{
\epsfig{figure=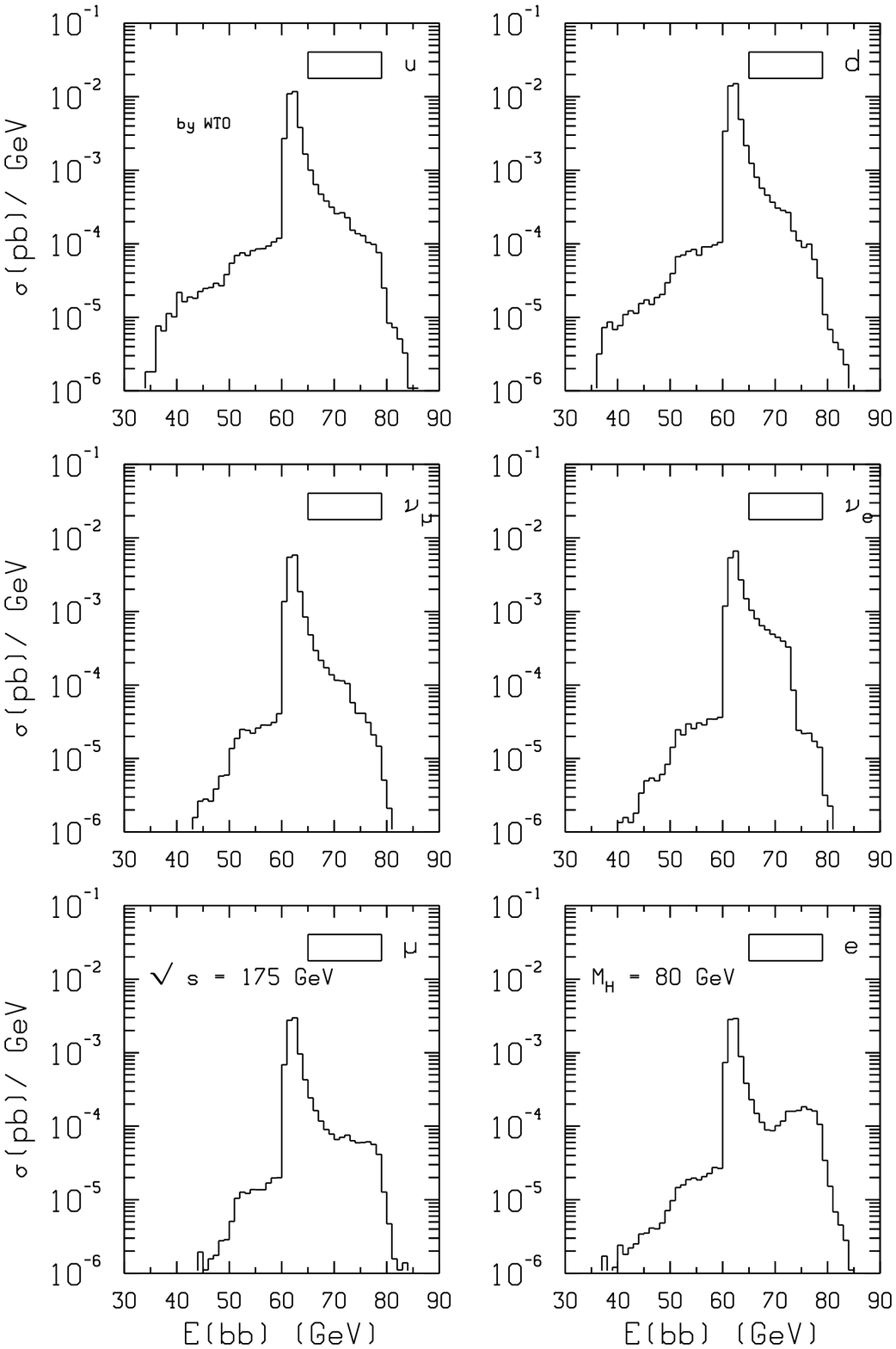,height=16cm,angle=0}
}
\vspace{0.1cm}
\begin{center}
{\bf Fig. 15}
\end{center}
\label{fgq}
\end{figure}

\newpage

\begin{figure}[htbp]
\vspace{0.1cm}
\centerline{
\epsfig{figure=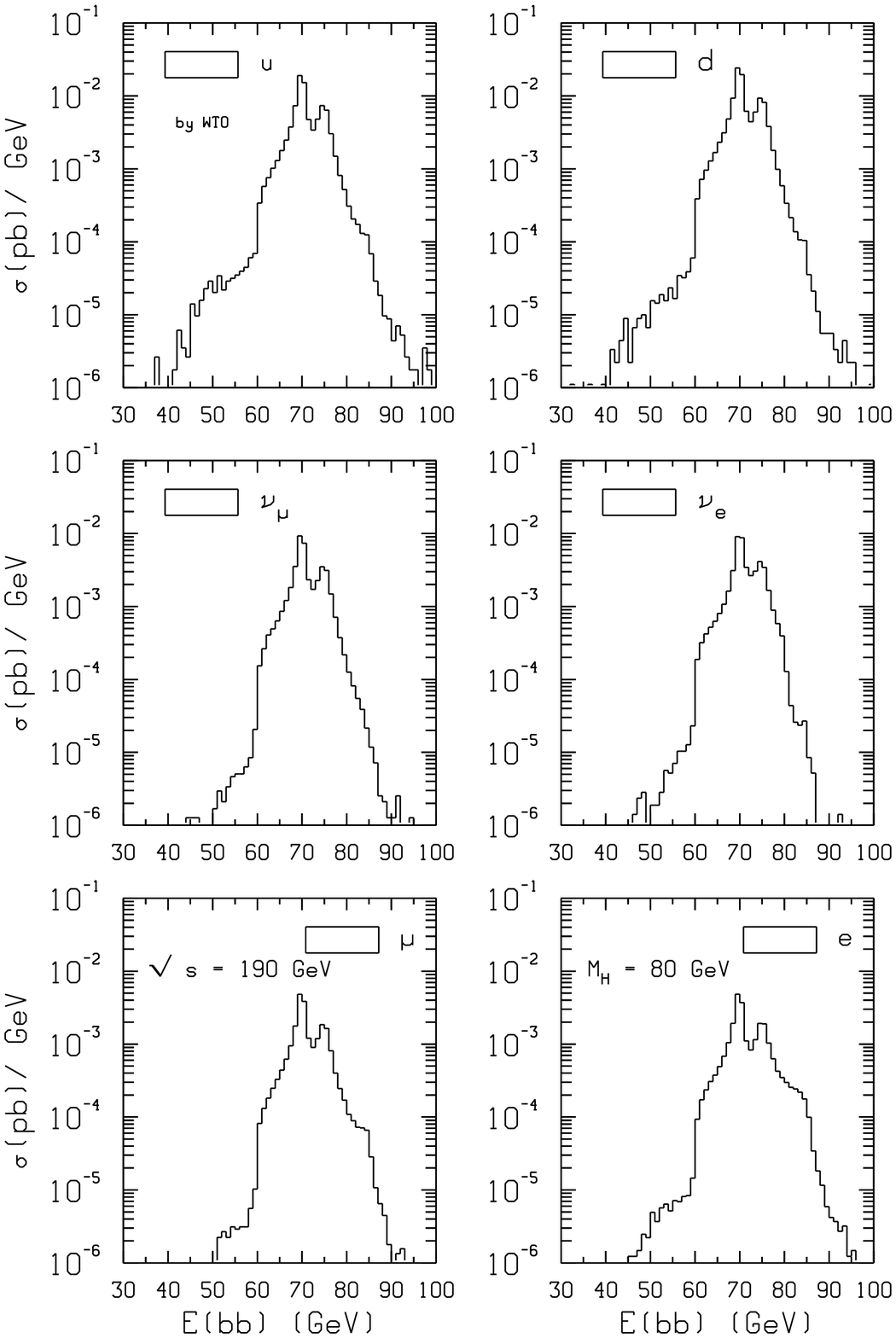,height=16cm,angle=0}
}
\vspace{0.1cm}
\begin{center}
{\bf Fig. 16}
\end{center}
\label{fgr}
\end{figure}

\newpage

\begin{figure}[htbp]
\vspace{0.1cm}
\centerline{
\epsfig{figure=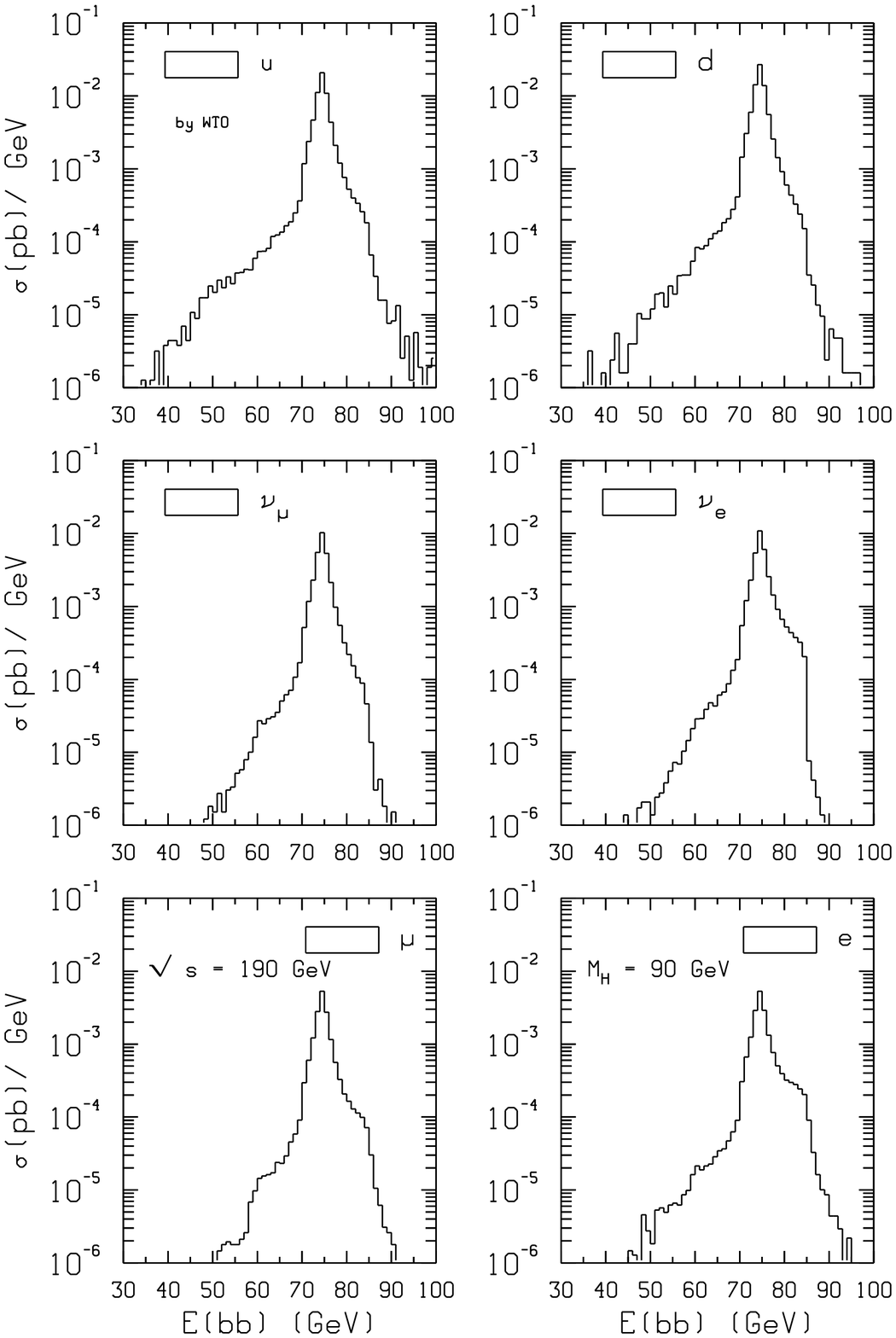,height=16cm,angle=0}
}
\vspace{0.1cm}
\begin{center}
{\bf Fig. 17}
\end{center}
\label{fgs}
\end{figure}

\newpage

\begin{figure}[htbp]
\vspace{0.1cm}
\centerline{
\epsfig{figure=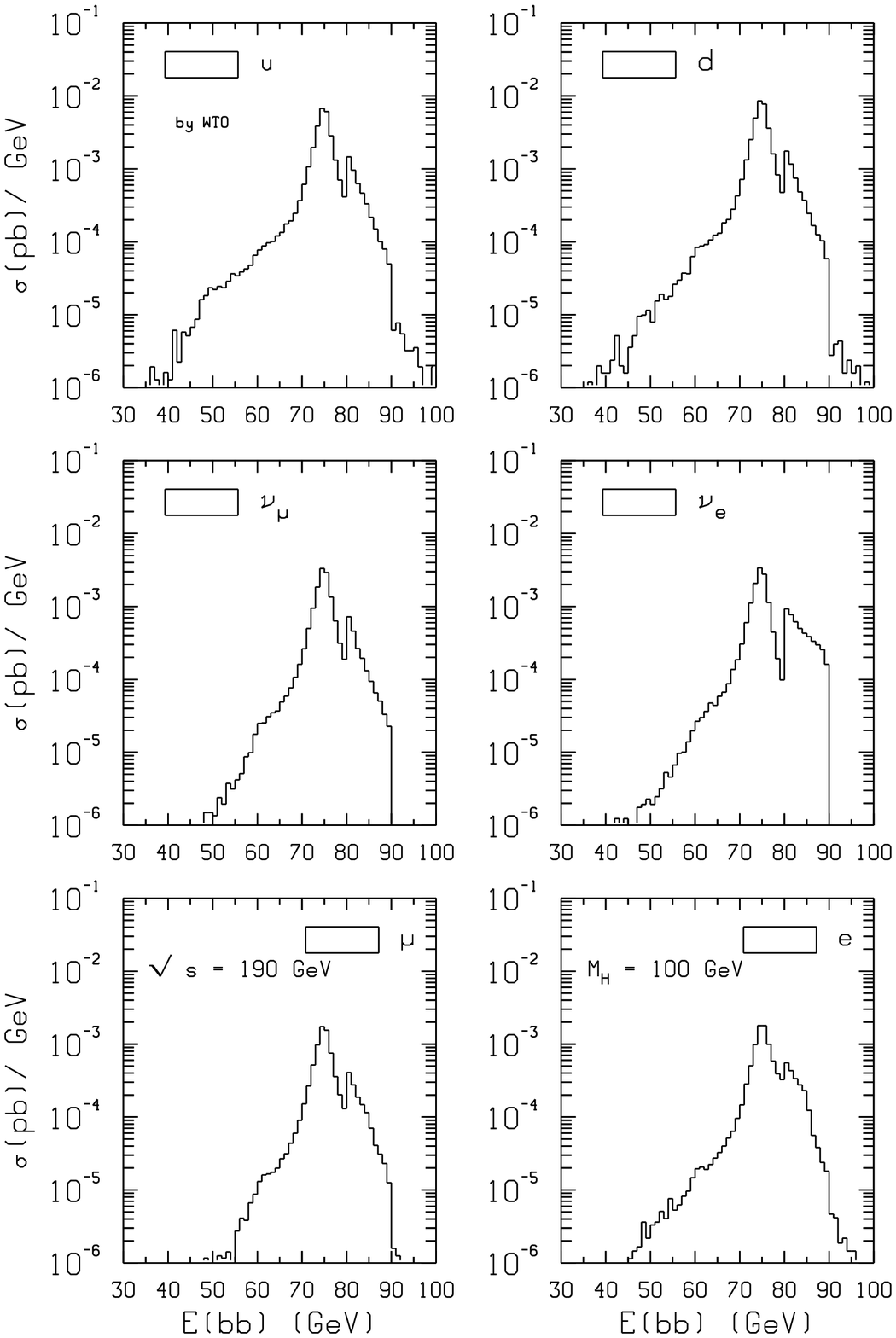,height=16cm,angle=0}
}
\vspace{0.1cm}
\begin{center}
{\bf Fig. 18}
\end{center}
\label{fgt}
\end{figure}

\newpage

\begin{figure}[htbp]
\vspace{0.1cm}
\centerline{
\epsfig{figure=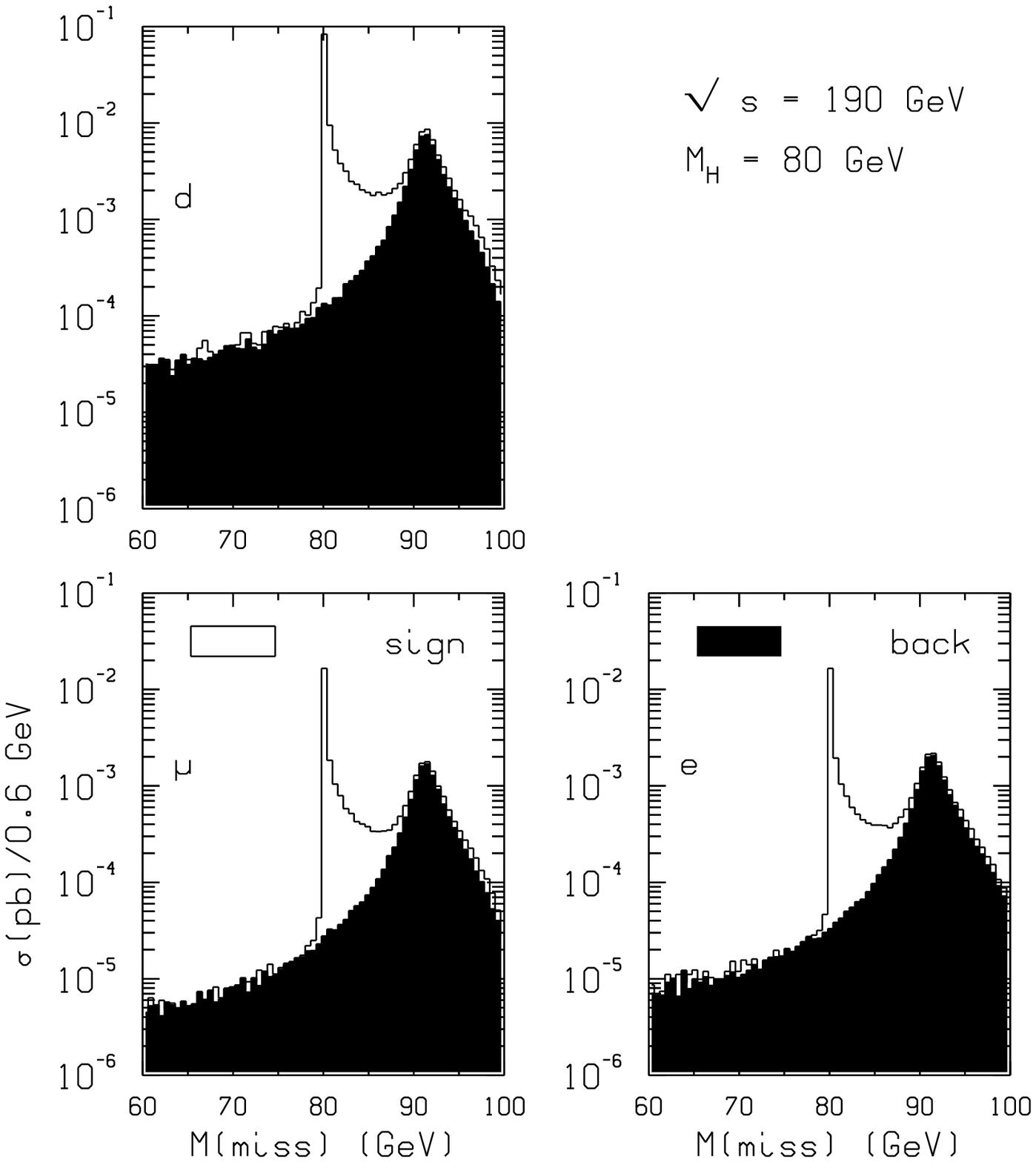,height=16cm,angle=0}
}
\vspace{0.1cm}
\begin{center}
{\bf Fig. 19}
\end{center}
\label{fgu}
\end{figure}

\newpage

\begin{figure}[htbp]
\vspace{0.1cm}
\centerline{
\epsfig{figure=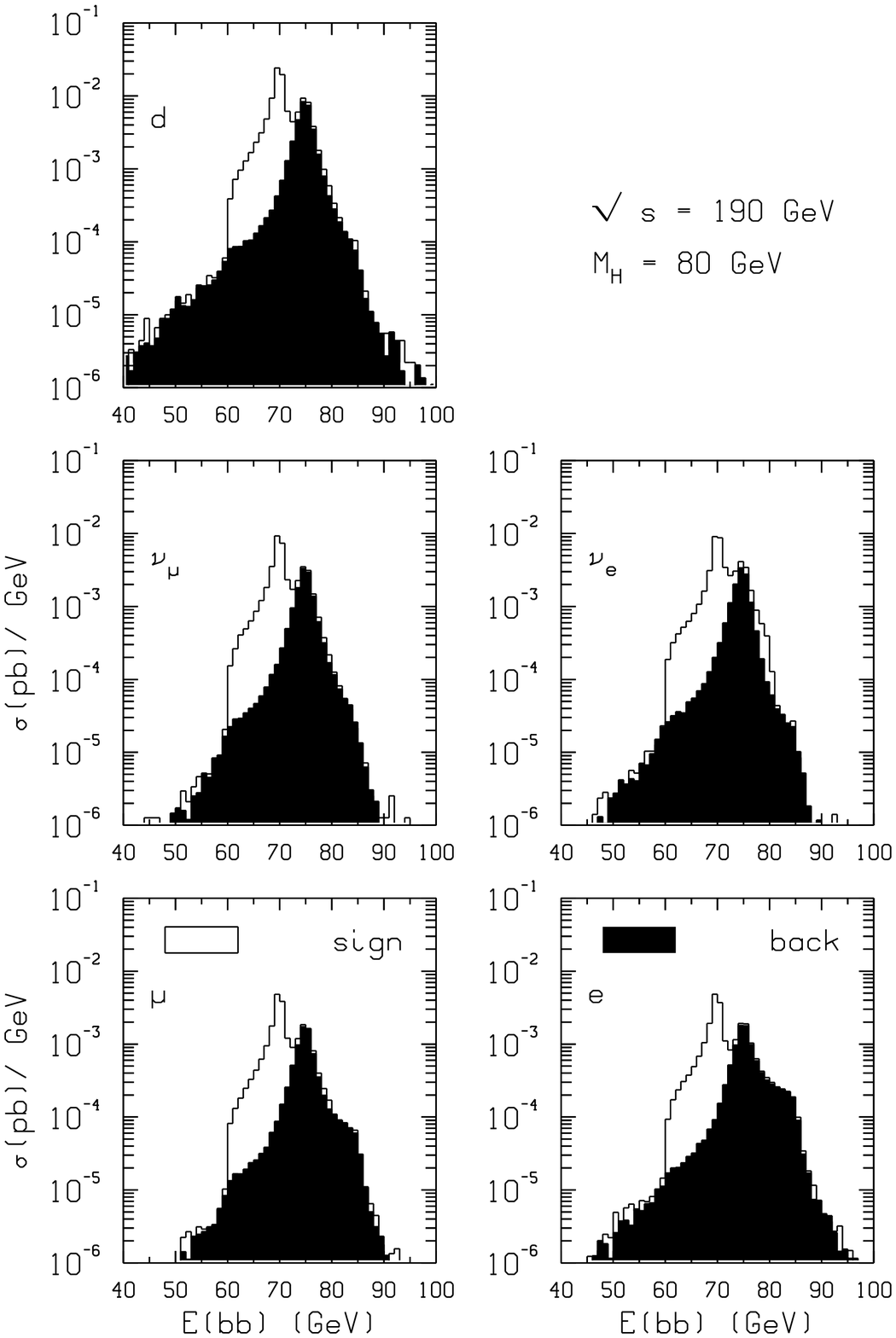,height=16cm,angle=0}
}
\vspace{0.1cm}
\begin{center}
{\bf Fig. 20}
\end{center}
\label{fgv}
\end{figure}

\newpage

\begin{figure}[htbp]
\vspace{0.1cm}
\centerline{
\epsfig{figure=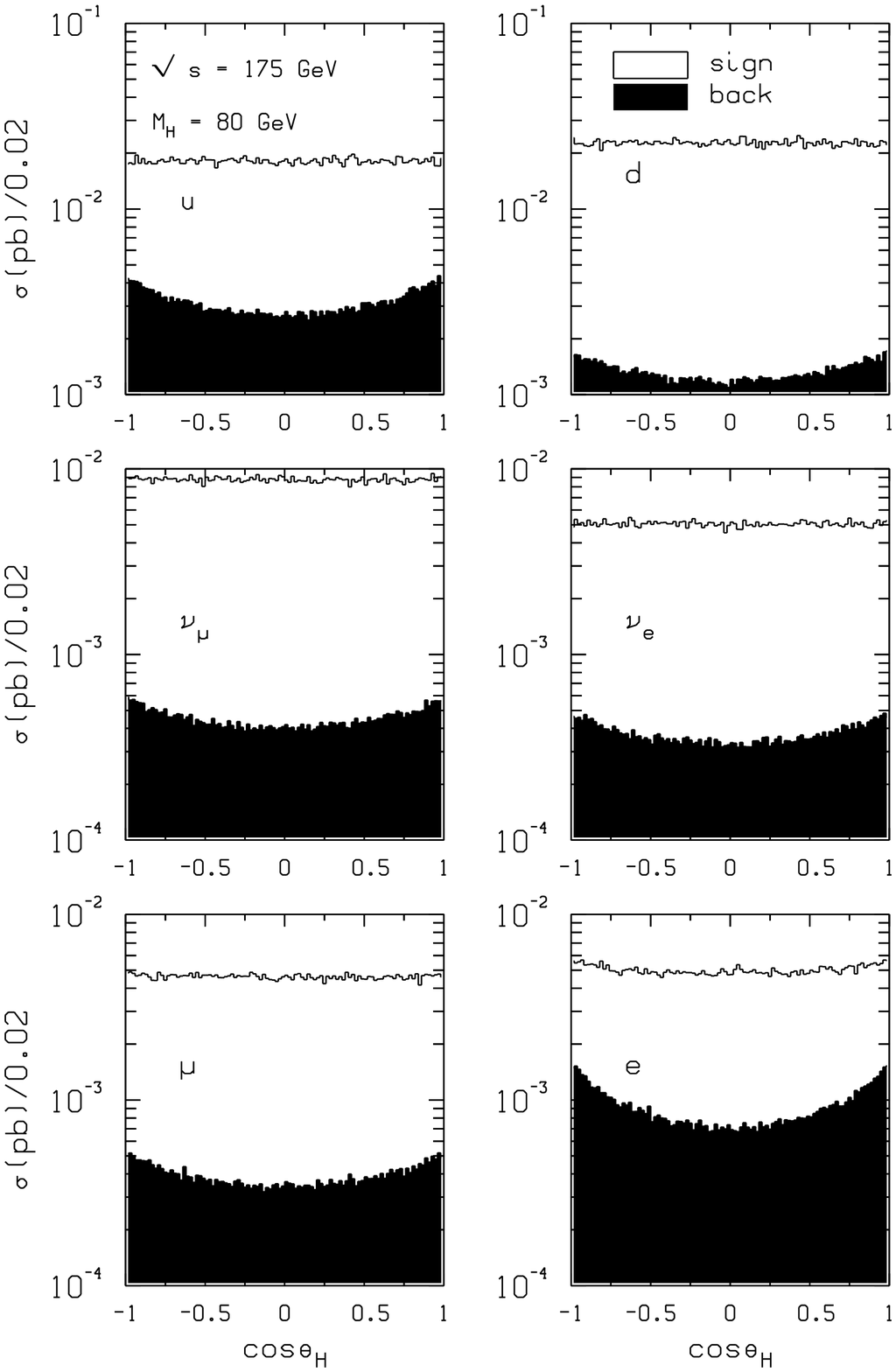,height=16cm,angle=0}
}
\vspace{0.1cm}
\begin{center}
{\bf Fig. 21}
\end{center}
\label{fgz}
\end{figure}


\begin{thebibliography}{99}

\bibitem{ichep} A.~Blondel, talk given at the 28th International Conference
on High Energy Physics, Warsaw 1996.

\bibitem{topaz0} G.~Montagna, O.~Nicrosini, G.~Passarino, F.~Piccinini and 
R.~Pittau, Comput. Phys. Commun. 76(1993)328.

\bibitem{exp} ALEPH Collaboration, CERN-PPE-79; \\
DELPHI Collaboration, CERN-PPE-119; \\
L3 Collaboration, CERN-PPE-95; \\
OPAL Collaboration, CERN-PPE-96.

\bibitem{yr2} {\it Event Generators for WW Physics}
to appear in Vol.1, {\it Report of the Workshop in Physics at 
LEP2} G.~Altarelli et. al., CERN-96-01; \\
{\it Event Generators for Discovery Physics}, M.~L.~Mangano et al.,
to appear in Vol.1, {\it Report of the Workshop in Physics at 
LEP2} G.~Altarelli et. al., CERN-96-01; \\
{\it Higgs Physics at LEP II}, M.~Carena et al.,
to appear in Vol.1, {\it Report of the Workshop in Physics at 
LEP2} G.~Altarelli et. al., CERN-96-01.

\bibitem{pc} A.~Ballestrero, private communication.

\bibitem{sh} A.~Djouadi et al., hep-ph/9605437;\\
B.~Kniehl, hep-ph/9610500.

\bibitem{up} W.~Beenakker et al. in preparation; \\
E.~N.~ Argyres et al., Phys. Lett. B358(1995)339; \\
R.~G.~Stuart, Phys. Lett. B262(1991)113;  \\
R.~G.~Stuart, Phys. Lett. B272(1991)353;  \\
Y.~Kurihara, D.~Perret-Gallix and Y.~Shimizu, Phys. Lett. B349(1995)367;  \\
U.~Baur and D.~Zeppenfeld, Phys. Rev. Lett. 75(1995)1002.

\bibitem{yr2ww} {\it WW Cross-Sections and Distributions}, W.~Beenakker
et al., to appear in Vol.1, {\it Report of the Workshop in Physics at 
LEP2} G.~Altarelli et. al., CERN-96-01.

\bibitem{tb} T.~Sj\"ostrand, report of the QCD generators group of the
LEP~2 workshop, to appear in Vol.1, {\it Report of the Workshop in Physics at 
LEP2} G.~Altarelli et. al., CERN-96-01.


\bibitem{jeg} F.~Jegerlehner, Z. Phys. C32(1986)195; \\
H.~Burkhardt et al., Z. Phys. C42(1989)497; \\
S.~Eidelman and F.~Jegerlehner, Z. Phys. C(1995).

\bibitem{sf} 
F.~A.~Berends, R.~Kleiss and R.~Pittau, Nucl. Phys. 
B426(1994)344; \\
G.~Montagna, O.~Nicrosini and F.~Piccinini, Comp. Phys. Commun.
90(1995)141; \\
G.~Montagna, O.~Nicrosini, G.~Passarino and F.~Piccinini,
Phys. Lett. B348(1995)178.

\bibitem{rqm} K.~G.~Chetyrkin et al. in {\it Reports of the Working Group on 
Precision Calculations for the $Z$ Resonance}, D.~Bardin W.~Hollik and
G.~passarino eds., CERN-95-03, p. 175.
 
\bibitem{bd} E.~Boos and M.~Dubinin, Phys. Lett. 308B(1993)147; \\
E.~Boos, M.~Sachwitz, H.~Schreiber and S.~Shichanin, Z. Physik C61(1994)675; \\
M.~Dubinin, V.~Edneral, Y.~Kurihara and Y.~Shimizu, Phys. Lett. B239(1994)379.

\bibitem{dz} D.~Bardin, M.~Bilenky, A.~Olchevski and T.~Riemann, Phys. Lett.
B308(1993)403; \\
D.~Bardin, A.~Leike and T.~Riemann, Nucl. Phys. B(Proc. Suppl.) 37B(1994)274.

\bibitem{ob1} P.~Grosse-Wiesmann, D.~Haidt and J.~Schreiber in {\it $e^+e^-$
Collisions at $500\,$GeV: the Physics Potential}, P.~M.~Zerwas ed., 
DESY 92-123A, p. 37.

\bibitem{ob2} P.~Janot in {\it $e^+e^-$
Collisions at $500\,$GeV: the Physics Potential}, P.~M.~Zerwas ed., 
DESY 92-123A, p. 107.

\bibitem{cod1} E.~Boos, M.~Dubinin, V.~Edneral, V.~Ilyin, A.~Kryukov,
A.~Pukhov and S.~Shichanin in {\it New Computing Techniques in Physics 
Research II}, ed. by P.~Perret-Gallix, World Scientific, Singapore, 1992, 
p. 665.

\bibitem{cod2} F.A.~Berends, R.~Kleiss and R.~Pittau,
Comp.Phys.Comm. 85(1995)437;\\
Nucl.Phys. B424(1994)308;\\ 
Nucl.Phys. B426(1994)344.

\bibitem{cod3} J.~Hilgart, R.~Kleiss and F.~Le~Diberder, Comput. Phys. Commun.
75(1993)191.

\bibitem{cod4} D.~Bardin, A.~Leike and T.~Riemann, Phys. Lett. B344(1995)383; \\
D.~Bardin, A.~Leike and T.~Riemann, Phys. Lett. B353(1995)513.

\bibitem{cod5} G.~Montagna, O.~Nicrosini and F.~Piccinini, Phys. Lett. 
B348(1995)496.

\bibitem{cod6} T.~Sj\'ostrand, Comp. Phys. Comm. 82(1994)74.

\bibitem{cod7} E.~Accomando and A.~Ballestrero, WPHACT 1.0, hep-ph/9607317,
to appear in Comp.~Phys.~Comm.

\bibitem{cod8} G.~Passarino,  Comp.~Phys.~Comm. 97(1996)261. 

\bibitem{to} T.~Ohl, Talk given at the 3rd International Symposium on Radiative 
Corrections, Cracow, Poland, August 1-5, 1996, to appear in the Proceedings.

\bibitem{kat} D.~Apostolakis, P.~Ditsas and S.~Katsanevas, hep-ph/9603383.

\bibitem{mhf} G.~Passarino, Nucl. Phys. B237(1984)249.

\bibitem{crad} G.~Passarino, Standard Model and Electroweak Interaction: 
Phenomenology, Talk given at the 3rd International Symposium on Radiative 
Corrections, Cracow, Poland, August 1-5, 1996, to appear in the Proceedings.

\bibitem{wto2} G.~Passarino, WTO 2.0 Event Generation and Full one loop 
fermionic corrections, in preparation.

\bibitem{susy} J.~F.~Gunion and H.~Haber, hep-ph/9610337; \\
M.~Quiros, hep-ph/9609392.

\end{thebibliography}
\end{document}